\documentclass[bibyear]{aa}  

\usepackage{graphicx}
\usepackage[utf8]{inputenc}
\usepackage{txfonts}
\usepackage{color}
\usepackage{xcolor}

\usepackage{amsfonts}
\usepackage{amsmath}
\usepackage{bm}
\usepackage{xcolor}
\usepackage{amssymb}
\usepackage{natbib}
\usepackage[normalem]{ulem}

\usepackage{cases}
\usepackage{orcidlink}

\begin{document}

   \title{Joint action of phase mixing and nonlinear effects in MHD waves propagating in coronal loops}


   \author{C. Meringolo \orcidlink{0000-0001-8694-3058} \inst{1,2}
          \and
          F. Pucci \orcidlink{0000-0002-5272-5404} \inst{3}
          \and
          G. Nisticò \orcidlink{0000-0003-2566-2820} \inst{1}
          \and
          O. Pezzi \orcidlink{0000-0002-7638-1706} \inst{3}
          \and
          S. Servidio \orcidlink{0000-0001-8184-2151} \inst{1,4}
          \and 
          F. Malara \orcidlink{0000-0002-5554-8765} \inst{1,4}
          }

   \institute{Dipartimento di Fisica, Università della Calabria, via P. Bucci, 87036 Rende (CS), Italy 
        \and
             Institut für Theoretische Physik, Max-von-Laue-Strasse 1, D-60438 Frankfurt, Germany
         \and
             Istituto per la Scienza e Tecnologia dei Plasmi, Consiglio Nazionale delle Ricerche (ISTP-CNR), 70126 Bari, Italy
         \and
             Istituto Nazionale di Astrofisica - INAF, Direzione Scientifica, Viale del Parco Mellini 84, 00136 Roma, Italy \\
        \email{claudiomeringolo@unical.it} }


 
  \abstract
  {The evolution of Alfv\'en waves in cylindrical magnetic flux tubes, which represent a basic model for loops observed in the solar corona, can be affected by phase mixing and turbulent cascade. Phase mixing results from transverse inhomogeneities in the Alfv\'en speed, causing different shells of the flux tube to oscillate at different frequencies, thus forming increasingly smaller spatial scales in the direction perpendicular to the guide field. Turbulent cascade also contributes to the dissipation of the bulk energy of the waves through the generation of smaller spatial scales. Both processes present characteristic time scales. Different regimes can be envisaged according to how those time scales are related and to the typical time scale at which dissipation is at work. }
   {We investigate the interplay of phase mixing and the nonlinear turbulent cascade in the evolution and dissipation of Alfv\'en waves using compressible magnetohydrodynamics numerical simulations. We consider perturbations in the form of torsional waves, both propagating and standing, or turbulent fluctuations, or a combination of the two. The main purpose is to study how phase mixing and nonlinear couplings jointly work to produce small scales in different regimes.}
   {We conduct a numerical campaign to explore the typical parameters as the loop length, the amplitude and spatial profile of the perturbations, and the dissipative coefficients. A pseudo-spectral code is employed to solve the three-dimensional compressible magnetohydrodynamic equations, modeling the evolution of perturbations propagating in a flux tube corresponding to an equilibrium configuration with cylindrical symmetry. }
    {We find that phase mixing takes place for moderate amplitudes of the turbulent component even in a distorted, non-axisymmetric configuration, building small scales that are locally transverse to the density gradient. The dissipative time decreases with increasing the percentage of the turbulent component. This behavior is verified both for propagating and standing waves. Even in the fully turbulent case, a mechanism qualitatively similar to phase mixing occurs: it actively generates small scales together with the nonlinear cascade, thus providing the shortest dissipative time. General considerations are given to identify this regime in the parameter space. The turbulent perturbation also distorts the background density, locally increasing the Alfv\'en velocity gradient and further contributing to accelerating the formation of small scales. 
    }
   {Our campaign of simulations is relevant for the coronal plasma where Reynolds and Lundquist numbers are extremely high. For sufficiently low perturbations' amplitude, phase mixing and turbulence work synergically, speeding up the dissipation of the perturbation energy: phase mixing dominates at early times and nonlinear effects at later times. We find that the dissipative time is shorter than those of phase mixing and the nonlinear cascade when individually considered.}
   

   \maketitle
%

\section{Introduction}

The dissipation of waves in the solar corona is considered one of the viable mechanisms that can explain the heating of the coronal plasma, a long-standing, not fully solved problem. Photospheric motions would generate such waves, which propagate upward to the corona along the magnetic field connecting different layers of the solar atmosphere, eventually dissipating within the corona.

There is much observational evidence of the presence of different wave modes in the corona, starting from early observations of nonthermal broadening of coronal lines \citep{Feldman88,Dere93}, detection of slow magnetohydrodynamic (MHD) waves \citep{Chae98,Ofman99} and of transverse oscillations in post-flare loops \citep{Nakariakov99,Schrijver99,Aschwanden99}. More recently, the excitation of standing kink oscillations in coronal loops has been studied \citep[e.g.,][]{Zimovets15,Goddard16} that could be excited by coronal impulsive events through a variety of mechanisms (see also \citet{Nakariakov21} for a recent review). While the above oscillations are subject to damping within a few wave periods, other oscillations have been detected mainly in quiet loops, with no apparent damping for many periods or even with temporary growing amplitude \citep{Wang12,Tian12,Nistico13}. Waveperiod scaling linearly with the loop length indicates that those decayless oscillations are standing waves \citep{Anfinogentov15}. Decayless oscillations could be driven by the interaction between loops and quasi-steady flows \citep{Nakariakov16}, or by continuous footpoint driving \citep{Karampelas17,Karampelas19,Guo19,Afanasyev20}. 
Moreover, it has been proposed that they could be related to Kelvin-Helmholtz (KH) rolls induced by transverse loop oscillations \citep{Antolin16}.
In addition, the ubiquitous presence of propagating, mainly-transverse waves has been revealed by both ground-based (CoMP, \citet{Tomczyk07,Tomczyk09}) and space-based (AIA, on board of SDO \citep{McIntosh11,Lemen12}) observations, with a dominance of outward to inward wave power \citep{Tomczyk09}. Those oscillations have been interpreted as Alfv\'en waves, though it has also been proposed that they could be fast-mode kink waves propagating in cylindrical structures \citep{VanDoorsselaere08}. Their ubiquity seems to indicate a possible role in coronal heating.

From a theoretical point of view, wave dissipation in the coronal plasma represents a non-trivial problem. The Reynolds and Lundquist numbers are estimated to be extremely large in the solar corona. This implies that a wave can be efficiently dissipated only if its energy is transferred to fluctuations at very small spatial scales. Such a process could require a very long dissipative time, which could be much longer than the coronal cooling timescale. The presence of inhomogeneities in the background structures can speed up the formation of small scales. In particular, when a transverse gradient of the Alfv\'en velocity is present, Alfv\'en waves are subject to phase mixing \citep{Heyvaerts83}. During phase mixing, wavefronts are bent due to differences in their propagation velocity on nearby field lines. This results in a progressive generation of increasingly small scales in the direction transverse to the background magnetic field. Phase mixing has been extensively studied both by following a normal-mode approach \citep{Steinolfson85,Califano90,Califano92} and by considering the evolution of an initial disturbance \citep{Lee86,Malara92,Malara96}. Effects of density stratification and magnetic line divergence have also been considered \citep{Ruderman98}, as well as nonlinear coupling with compressive modes \citep{Nakariakov97,Nakariakov98}. Phase mixing in 3D configurations in the small wavelength limit has been studied \citep{Petkaki98,Malara00}, also in simplified models of quiet-Sun \citep{Malara03,Malara05,Malara07} or open fieldline \citep{Malara13,Pucci14} regions. It has also been shown that phase mixing is active in loops with a radial density inhomogeneity when azimuthally-polarized Alfv\'en waves are excited by coupling with kink modes \citep{Pagano17}. 

Torsional Alfv\'en waves represent one possible oscillating mode in a cylindrically symmetric structure with an axial magnetic field; they represent particular azimuthally polarized waves where fluctuations do not depend on the azimuthal angle $\theta$. They are non-compressive and are analogous to shear Alfv\'en waves in slab geometry. Their evolution is only due to phase mixing, except for large amplitudes when nonlinear effects come into play. Therefore, in studying phase mixing, torsional Alfv\'en waves represent the most suitable case to be considered. Such waves could be excited by torsional motions at the loop bases. Indeed, this kind of motion has been revealed in the lower layers of the solar atmosphere. In the photosphere, vortical motions seem to be related to convection \citep{Brant88}, mainly located at the downdrafts where the plasma returns to the solar interior after cooling down \citep{Bonet08,Bonet10}. Small-scale swirl events have been revealed in the quiet-Sun chromosphere \citep{Wedemeyer09,Tziotziou18}, as well as rapidly-rotating magnetic structures, ubiquitously distributed in the transition region \citep{Wedemeyer12}. Automated detection of chromospheric swirls has been recently performed \citep{Dakanalis22}, finding a mean lifetime $3.4$ min with typical diameter $1$ to $1.5$ Mm and about $40 \%$ of the analyzed surface covered by swirls. Torsional waves have been detected in magnetic funnels, with an oscillation period of $126-700$ s \citep{Jess09}. An investigation of a small-scale chromospheric tornado \citep{Tziotziou20} has suggested the existence of waves propagating upwards with phase speeds of $\sim 20-30$ km/s, in the form of both kink modes and localized Alfv\'enic torsional waves. Numerical simulations of torsional waves in the magnetic field of a chromospheric funnel have been performed to study frequency filtering \citep{Fedun11}. Moreover, the presence of Alfv\'en waves has been detected in simulations of photospheric vortexes \citep{Shelyag13}.

Beside phase mixing, turbulence represents an alternative way of generating small scales in a magnetofluid. Turbulence in the solar wind has been characterized using a huge number of in-situ measurements for many decades (see, e.g., \citet{Bruno13} for a review). Recent direct measurements of turbulence have been performed by Parker Solar Probe spacecraft in the most external solar corona where the wind is still sub-sonic/sub-Alfv\'enic \citep{Kasper21,Bandyopadhyay22,Zhao22}. Moreover, indirect indications exist of turbulent fluctuations in the corona. In particular, the non-thermal broadening of coronal spectral lines could be indicative of turbulent fluctuations \citep{Banerjee98,Singh06,Hahn13,Hahn14}, as well as a $f^{-1}$ frequency spectrum found in CoMP observations of loops \citep{Morton16,Morton19}. At variance with phase mixing where 
a mode couples with the inhomogeneity of the background structure, in turbulence the generation of small scales is due to nonlinear coupling between fluctuations. This typically generates power-law spectra in the wave number space. Similar to phase mixing, the magnetohydrodynamic (MHD) turbulent cascade preferentially generates small scales in the direction perpendicular to the background magnetic field. This especially holds in the case of a strong magnetic field (low plasma $\beta$), as in the case of the coronal plasma \citep{shebalin83}. In Alfv\'enic turbulence, nonlinear couplings occur between fluctuations propagating in opposite senses. This situation can be easily envisioned in closed magnetic structures like loops. However, even in open-fieldline regions, partial wave reflection due to longitudinal inhomogeneities can give rise to opposite-propagating fluctuations and a consequent activation of a nonlinear cascade. Models of coronal heating based on turbulence have been formulated for coronal loops \citep{vanBallegooijen11,Downs16,vanBallegooijen17,vanBallegooijen18, rappazzo2017}, as well as for open structures \citep{Verdini09,Perez13,Woolsey15,Chandran19}. Hybrid shell models are simplified turbulence models based on reduced MHD and hold for a low-$\beta$ plasma dominated by nearly transverse non-compressive fluctuations. They have been applied to the heating of coronal loops \citep{Nigro04,Nigro05,Reale05,Buchlin07}, as well as to characterize the nonlinear energy spectral flux \citep{Nigro08a,Malara10}. Such models reproduce power-law frequency spectra \citep{Nigro20} similar to those observed in the corona \citep{Tomczyk09,McIntosh11,Morton16}  

A situation where both phase mixing and turbulence are active is when an initial standing kink mode resonantly couples with Alfv\'enic oscillations located at the interface between the loop interior and exterior. The spatial variation of the Alfv\'en velocity at the interface generates phase mixing in the Alfv\'enic azimuthal oscillation, giving origin to a KH instability. The final result is a turbulent state in the interface region that eventually leads to wave energy dissipation. This scenario has been studied in great detail \citep{Terradas08,Antolin14,Magyar15,Antolin19}, also in comparison with observations \citep{Antolin17}, pointing out the role played by KH rolls in enhancing dissipation, as well as in the formation of fine strand-like structures \citep{Antolin14}.

Finally, it has been shown that in a plasma with transverse inhomogeneities phase mixing can result in a phenomenology resembling that of turbulence also in the case in which uni-directional waves are present \citep{magyar2017generalized,magyar2019understanding}.

The present paper is devoted to studying the interplay between phase mixing and turbulence in the generation of small scales and consequent wave dissipation in a simple model of a coronal loop. Since we are interested in selecting the effects of phase mixing, we will consider torsional modes with a possible superposition of turbulent perturbation here, excluding other more complex situations such as kink modes. We will show that the simultaneous presence of phase mixing and turbulent cascade can act in a synergistic way, such as to reinforce both effects. This can lead to an efficient dissipation, even in situations where small amplitude waves propagate in weakly dissipative plasma, as for the solar corona. 

The plan of the paper is the following.
An overview of the model is presented in Sect. \ref{sec2}, where we report some relevant scaling laws in Sect. \ref{sec2.1} and the physical model we aim to describe in Sects. \ref{sec2.3}--\ref{turb}. 
In Sect. \ref{sec3}, we report our numerical results. We first focus on the purely Alfvén wave perturbations in Sect. \ref{sec3.1} and on the turbulence in a homogeneous background in Sect. \ref{append}, moving to the combined case in Sect. \ref{sec3.3}, both for propagating and standing waves.
Finally, in Sect. \ref{sec4} we discuss our results and its implications.

\section{Model}
\label{sec2}

We aim to model a situation where transverse fluctuations propagate and evolve inside a coronal loop characterized by a transverse inhomogeneity. We are interested in studying how the coupling between background inhomogeneity and fluctuations and nonlinear couplings among fluctuations lead to the formation of small-scale fluctuations and eventually to their dissipation. As described below, the loop is modeled as a straight magnetic flux tube with a cylindrical symmetry (at the initial time) where the density in the inner part of the loop is larger than in the outer part, while the magnetic field is axially directed and nearly uniform, due to a low plasma-$\beta$. This results in a transverse modulation of the Alfv\'en velocity $c_A(r)=B(r)/\sqrt{4\pi \rho(r)}$, where $r$ is the radial coordinate with respect to the loop axis, as it can be appreciated in Fig. \ref{icf}. Different kinds of fluctuations will be considered, namely, (i) torsional Alfv\'en waves, (ii) turbulent transverse fluctuations, and (iii) a superposition of (i) and (ii). 

\subsection{Phenomenology and scaling laws}
\label{sec2.1}

Prior to describing the model and its results, we discuss some general results concerning wave dynamics in the considered situation and their implication for our model. 

When an Alfv\'enic fluctuation propagates in a plasma with a transverse inhomogeneous Alfv\'en velocity, it is subject to phase mixing that progressively generates small scales perpendicularly to the background magnetic field $\bm{B}_0$ \citep[e.g.,][]{Heyvaerts83}. We indicate by $k_\perp$ and $k_{||}$ the wavevector components perpendicular and parallel to $\bm{B}_0$, respectively. During phase mixing $k_{||}$ remains constant: $k_{||}=k_{||0}$, while $k_\perp$ linearly increases with time $t$, according to:
\begin{equation}\label{kpm}
k_\perp (t) \sim k_{||0} \frac{dc_A}{dr} t
\end{equation}
where $dc_A/dr$ is a typical value for the transverse gradient of the Alfv\'en velocity. For simplicity, in Eq. (\ref{kpm}) we have assumed that $k_\perp(t=0)=0$. 
Following Eq.~\ref{kpm}, we define the phase-mixing dynamical time $t_{PM}(k_\perp)$ as 
\begin{equation}\label{tPM0}
t_{PM}(k_\perp) \sim \frac{k_\perp}{k_{||0}} \left( \frac{dc_A}{dr} \right)^{-1}
\end{equation}
which is also the time that is needed for a given perpendicular wavevector $k_\perp$ to double.
The stronger the transverse gradient of the Alfv\'en speed, the faster phase mixing. Moreover, $t_{PM}$ increases with increasing $k_\perp$. Therefore, as a mechanism able to generate small scales, phase mixing becomes progressively less efficient with increasing the fluctuation wavevector. 

Indicating by $\ell_d \sim k_{\perp d}^{-1}$ the dissipation length and with $k_{\perp d}$ the corresponding perpendicular wavevector, we define the dissipative time $t_d^{PM}$ as the time phase mixing takes to increase $k_\perp(t)$ up to $k_{\perp d}$. Using Eq. (\ref{kpm}) we obtain:
\begin{equation}\label{tdPM0}
t_d^{PM} \sim \frac{k_{\perp d}}{k_{||0}} \left( \frac{dc_A}{dr}\right)^{-1} \sim \frac{L_{||}}{\ell_d} \left( \frac{dc_A}{dr}\right)^{-1}
\end{equation}
where we have assumed that the initial parallel wavelength is of the order of the loop length $L_{||}$. The above expression can be normalized to a typical time scale $t_A=L_\perp/c_{A0}$, where $L_\perp$ is the loop width and $c_{A0}$ is the mean Alfv\'en velocity. We obtain:
\begin{equation}\label{ratio1}
\frac{t_d^{PM}}{t_A} \sim \frac{c_{A0}}{\ell_d} \left( \frac{dc_A}{dr}\right)^{-1} \frac{L_{||}}{L_\perp}
\end{equation}
where $\Lambda=L_{||}/L_\perp$ is the loop aspect ratio. To give an order-of-magnitude estimation for the ratio (\ref{ratio1}, we assume that the density inside the loop is a factor 2 larger than outside and the magnetic field is nearly uniform as set in our numerical model (see Fig. \ref{icf}). Therefore, the Alfv\'en velocity has a relative variation of about $25\%$ from inside to outside the loop. We also assume that $c_A$ varies over a scale $\Delta r \sim L_\perp/10$, with $L_\perp \sim 10^3$ km and a loop aspect ratio $L_{||}/L_\perp \sim 30$. In the coronal plasma, the Lundquist number is extremely high and dissipation is probably due to kinetic effects; therefore, we consider a dissipation length of the order of proton Larmor radius $\ell_d \sim 10^{-3}$ km. Using those values, we obtain an estimation for the normalized phase-mixing dissipative time: $t_d^{PM}/t_A \sim 10^7$. Such a large value indicates that phase mixing alone is inefficient in dissipating Alfv\'en waves, at least in the simple situation here considered.

Another possible dissipative mechanism is represented by a turbulent cascade, where fluctuating energy is transferred to smaller scales by nonlinear couplings. In MHD, the turbulent cascade mainly takes place perpendicular to the background magnetic field \citep[e.g.,][]{shebalin83}, and such effect is even stronger in the low-$\beta$ coronal plasma. 
Since the turbulent cascade can virtually generate indefinitely small scales in a finite time, the dissipative time is larger but of the order of the nonlinear time $t_{NL}(\ell_{\perp 0})$, where $\ell_{\perp 0}$ is the large energy-containing scale of fluctuations. Namely: 
\begin{equation}\label{tNLMHD}
t_d^{\text{turb}} \sim \Gamma t_{NL}(\ell_{\perp 0}) = \Gamma \frac{\ell_{\perp 0}}{c_{A0}}\frac{B_0}{\delta B(\ell_{\perp 0})}
\end{equation}
where $\delta B(\ell_{\perp 0})$ is the amplitude of magnetic field fluctuation at the scale $\ell_{\perp 0}$, $c_{A0}$ is the mean Alfv\'en speed and $\Gamma $ is a numerical factor of the order of a few units. 
Normalized to $t_A$, it is:
\begin{equation}\label{tdturb0}
\frac{t_d^{\rm turb}}{t_A} \sim \Gamma \frac{\ell_{\perp 0}}{L_\perp} \frac{B_0}{\delta B(\ell_{\perp 0})}
\end{equation}
We notice that the dissipative time increases with decreasing fluctuations' amplitude $\delta B$.
Assuming that the largest scale in fluctuation is of the order of the loop width: $\ell_{\perp 0} \sim L_\perp$ and a normalized fluctuation amplitude $\delta B/B_0 \sim 10^{-2}$, we obtain for the normalized dissipative time: $t_d^{\rm turb}/t_A \sim 100 \Gamma$. This value is many orders of magnitude smaller than the above estimation made for the phase-mixing case. Therefore, in a high Reynolds/Lundquist plasma, as in the corona, turbulence is more efficient than phase mixing to dissipate fluctuations, even for small-amplitude fluctuations.

However, the above picture could change in a more realistic case where turbulence evolves inside an inhomogeneous background, such as a coronal loop. In such a case, besides nonlinear effects driving the turbulent cascade, the coupling between the background inhomogeneity and fluctuations (that is at the bases of phase mixing) also contributes to generating small scales. Therefore, it can be expected that in such a configuration phase mixing and nonlinear effects work together to promote small-scale generation and speed-up dissipation. In particular, let us imagine a situation where, at the initial time, the fluctuation amplitude at large scale $\delta B(\ell_{\perp 0})$ is small enough to have $t_{PM}(\ell_{\perp 0}) <  t_d^{\text{turb}}$.
Using Eq.s (\ref{tPM0}) and (\ref{tNLMHD}) this condition can be expressed by:
\begin{equation}\label{cond}
    \dfrac{\delta B}{B_0} < \dfrac{dc_A}{dr} \dfrac{\ell^2_{\perp 0}}{L_{\parallel}}   \dfrac{\Gamma}{c_{A0}} .
\end{equation}
where $\delta B \equiv \delta B (\ell_{\perp0})$.
In such a case, at an early stage of the time evolution phase, mixing would proceed faster than nonlinear effects to generate small scales. Moreover, phase mixing leaves the fluctuation amplitude $\delta B(\ell_\perp)$ unchanged as $\ell_\perp$ decreases. Therefore, as $\ell_\perp$ decreases in time, $t_{NL}(\ell_\perp)$ will proportionally decrease, too. Hence, at a time $t^*$ and at a corresponding scale $\ell_\perp^*$, the situation will be reversed, namely, $t_d^{\text{turb}}(\ell_\perp) < t_{PM}(\ell_\perp)$ for $\ell_\perp < \ell_\perp^*$. From that time on, nonlinear effects become the faster mechanism producing small scales and the cascade will bring energy to dissipative scales in a time $\gtrsim t_d^{\text{turb}}(\ell_\perp^*)$, which is smaller than the initial $t_d^{\text{turb}}$. In this scenario, the two considered mechanisms work in a synergistic way: phase mixing dominates at early times ($t\lesssim t^*$) and nonlinear cascade at later times ($t\gtrsim t^*$). The result is a dissipative time that is shorter than those of phase mixing and of the nonlinear cascade when individually considered. 

In what follows, we will explore the above ideas from a quantitative point of view by means of numerical simulations.


\subsection{MHD equations and the numerical method}
\label{sec2.2}

To describe the evolution of perturbations in the coronal plasma we use the compressible MHD equations that are written in the following form, using dimensionless quantities: 
\begin{eqnarray}
\frac{\partial \rho}{\partial t} &=& - \mathbf{\nabla} \cdot \left( \rho {\bm v} \right),\label{MHD1}
\\
\frac{\partial \bm{v}}{\partial t} &=& - \left( \bm{v} \cdot \bm{\nabla} \right) \bm{v} + \frac{1}{\rho} \left[ \left(\bm{\nabla} \times \bm{B}\right)\times\bm{B} \right] -\nonumber 
\\
&&  \frac{{\beta}}{2\rho}\bm{\nabla}\left(\rho T\right)  -\nu_4 \bm{\nabla}^4 \bm{v},\label{MHD2} 
\\
\frac{\partial \bm{A}}{\partial t} &=&  \bm{v} \times \bm{B} -\eta_4 \bm{\nabla}^4 \bm{A} ,\label{MHD3} 
\\
\frac{\partial T}{\partial t} &=& - \left( \bm{v} \cdot \bm{\nabla} \right) T - \left( \gamma -1 \right) \left( \bm{\nabla} \cdot \bm{v} \right) T -\chi_4 \bm{\nabla}^4 T.
\label{MHD4}
\end{eqnarray}
Here, $\rho$ is the density normalized to a typical value $\Tilde{\rho}$; $\bm{B}$ is the magnetic field 
normalized to a typical value $\Tilde{B}$; $\bm{A}$ is the vector potential normalized to $\Tilde{\ell} \Tilde{B}$, with $\Tilde{\ell}$ a typical length; 
$\bm{A}$ is related to $\bm{B}$ through $\bm{B} = \bm{\nabla} \times \bm{A}$;  
$\bm{v}$ is the magnetofluid velocity normalized to the typical Alfv\'en velocity $\Tilde{c}_A=\Tilde{B}/\left[ 
(4 \pi \Tilde{\rho})^{1/2}\right]$; $T$ is the temperature normalized to a typical value $\Tilde{T}$; $\gamma$ is
the adiabatic index. Spatial coordinates $x$, $y$ and $z$ are normalized to $\Tilde{\ell}$ and 
time is normalized to the Alfv\'en time $\Tilde{\ell} /\Tilde{c}_A$. 
The coefficient ${\beta}$ is the plasma beta, defined as the ratio between the typical kinetic and the magnetic pressure, 
${\beta}=8\pi \kappa_B \Tilde{\rho} \Tilde{T}/(\mu m_p \Tilde{B}^2)$, with $\kappa_B$ the Boltzmann constant, 
$\mu$ the mean atomic weight and $m_p$ the proton mass. In the coronal plasma, it is typically ${\beta} \ll 1$; we 
used the value ${\beta}=0.05$. In code units, the plasma pressure is given by $P=\beta \rho T/2$.

Compressibility is included in the model. Indeed, as discussed, for instance, in \citep{Malara96}, the production of small scales necessary for the occurrence of dissipation is more effective in a compressible medium rather than in an incompressible one. The energy equation (\ref{MHD4}) expresses the adiabaticity condition (except for the thermal hyper-diffusivity, see below). In fact, since we are interested in describing the route to energy dissipation rather than dissipation itself, we use the simplified energy equation (\ref{MHD4}) where effects such as radiative losses are neglected.
Integrating Eq. (\ref{MHD3}), which describes the time evolution of the vector potential $\bm{A}$ instead of the magnetic field $\bm{B}$ guarantees $\nabla \cdot \bm{B} = 0$.
Moreover, a logarithmic regularization to the density is used; namely, we set $\rho = e^{g}$, and solve the equivalent equation for $g$ with the purpose of a better description of possible discontinuities and shocks. 
Hyper-dissipation is implemented through the last terms in Eq.s (\ref{MHD2})--(\ref{MHD4}) representing hyper-viscosity, hyper-resistivity 
and hyper thermal diffusivity, respectively, with $\nu_4$, $\eta_4$ and $\chi_4$ the corresponding 
coefficients. These terms have been introduced to dissipate the turbulent cascade at small scales and 
control numerical stability; they do not mimic any physical process. With respect to standard dissipation, hyper-dissipation 
allows one to obtain more extended fluctuation spectra, keeping numerical stability. 

Eq.s (\ref{MHD1})--(\ref{MHD4}) are solved in a 3D Cartesian domain with periodicity boundary conditions along the three space directions $x$, $y$, and $z$. In particular, $z$ represents the direction of the background magnetic field $\bm{B}_0$.
The 3D Cartesian domain $D$ in normalized units is defined as 
$$D=\{x,y,z\}=[0 : L] \times [0 : L] \times [0 : \Lambda L],$$
with $L=2 \pi$.
Throughout the paper, we will indicate the domain size perpendicular to $\bm{B}_0$ by $L$, and the parallel size by $\Lambda L=L_\parallel$, where $\Lambda = L_\parallel / L$ is a free parameter that represents the domain aspect ratio.

To solve Eq.s (\ref{MHD1})-(\ref{MHD4}) in 3D configurations, we used the ``COmpressible Hall Magnetohydrodynamics simulator for Plasma Astrophysics'' (\texttt{COHMPA}) algorithm with the Hall term switched off.
A pseudo-spectral algorithm is used, which adopts a second-order Runge-Kutta scheme to advance in time the MHD variables.
In the physical domain $D$ all quantities are calculated on a regular grid formed by $N_\perp \times N_\perp \times N_{||}$ grid points: $\left( x_i, y_j, z_l \right) = \left(iL/N_\perp,jL/N_\perp,l\Lambda L/N_{||} \right)$, with $i,j,l$ integers, $0\le i,j \le N_\perp-1$ and $0 \le l \le N_{||}-1$. In all numerical runs, we used $N_\perp \ge N_{||}$. Correspondingly, a discrete set of wavevectors is defined in the spectral space: $\bm{k} = \left( 2\pi n_x/L, 2\pi n_y/L, 2\pi n_z/(\Lambda L) \right)$, with $n_x,n_y,n_z$ integers, $0\le n_x,n_y \le N_\perp/2-1$ and $0 \le n_z \le N_{||}/2-1$.
A spectral filter is employed to suppress numerical artefacts due to aliasing \citep{Orszag1971, Meringolo2021}, and a \textit{2/3 rule} is adopted where the spectral coefficients of all quantities are set to
zero at $k = |\bm{k}| > 2/3 k_{\rm N}$, with $k_{\rm N}=(2\pi/L)(N_\perp/2-1)$ corresponding to the Nyquist frequency \citep{Canuto1988}. A 2.5D version of this algorithm has already been adopted in literature \citep{vasconez2015, perri2017, pezzi2017revisiting}. More recently, a fully 3D version of the algorithm has been exploited by \citet{pezzi2023turbulence}.

The initial condition is given by the superposition of an ideal (non-dissipative) MHD equilibrium and a perturbation. The equilibrium represents a simplified model for a coronal loop. The perturbation represents a torsional propagating or standing Alfv\'en wave, or a turbulent perturbation, or a combination of both. The explicit forms of the equilibrium and the different kinds of perturbation are given in the following.

\subsection{Equilibrium: a cylindrical magnetic flux tube}
\label{sec2.3}

The equilibrium is a structure with cylindrical symmetry around an axis parallel to the $z$-axis. All quantities relative to the equilibrium are indicated by the index "0". The magnetic field $\bm{B}_0=B_0(x,y)\hat{{\bm z}}$ is directed along $z$, where $\hat{{\bm z}}$ is the unit vector in the $z$ direction. Magnetic lines have no curvature, and the Lorenz force per volume unit reduces to the gradient of magnetic pressure. Therefore, the equilibrium is sustained by the balance between magnetic and plasma pressure, expressed by the following relation
\begin{equation}
\label{equi}
\frac{B_0^2(x,y)}{2}+P_0(x,y)={\mbox{const}},
\end{equation} 
where $P_0(x,y)=\beta\rho_0(x,y)T_0/2$ is the equilibrium plasma pressure in dimensionless code units, $\rho_0(x,y)$ is the inhomogeneous density and $T_0$ is the temperature that is assumed to be uniform in space.
The inhomogeneities of equilibrium are transverse to the guide field $\bm{B}_0$. We have chosen the following functional form for the plasma pressure

\begin{equation}\label{press0}
P_0(x,y)=\frac{\left( P_{int}-P_{ext} \right)}{2} \left[ 1 - \tanh \left( \frac{r-r_0}{\Delta r} \right) \right] + P_{ext}.
\end{equation}
Here and in what follows, subscripts {\it int} and {\it ext} indicate values of quantities in regions internal or external to the flux tube, respectively. Moreover, $(x_0,y_0)=(L/2, L/2) = (\pi, \pi)$ is the position of the symmetry axis, $r=\sqrt{(x-x_0)^2+(y-y_0)^2}$ is the distance of a point from the symmetry axis, $r_0=L/4=\pi/2$ is the radius of the flux tube and $\Delta r=L/16=\pi/8$ is a parameter controlling the width of the shear region separating the interior from the exterior of the flux tube.

Equation (\ref{press0}) describes a cylindrically symmetric function. On the other hand, the domain $D$ is a parallelepiped with periodicity boundary conditions. Even though the expression (\ref{press0}) for $P_0(x,y)$ is periodic on the boundaries of $D$, its derivatives normal to the domain boundaries are not periodic, which could provoke numerical problems. 
Therefore, we have modified the expression for $P_0(x,y)$ so that all its first-order partial derivatives are periodic. Indicating the original form of $P_0$ (Eq. (\ref{press0})) by $P_{old}$ and the corrected form by $P_{corr}$, the two are related by:
\begin{equation}\label{Pcorr}
P_{corr}(x,y) = P_{old}(x,y) + a(y)\left(x-\frac{L}{2}\right)^2 + b(x)\left(y-\frac{L}{2}\right)^2.
\end{equation}
The functions $a(y)$ and $b(x)$ are chosen such as the first-order normal derivatives of $P_{corr}$ are vanishing all along the domain boundaries.
This condition is verified by the following choice:
\begin{equation}\label{a(y)}
a(y) = \frac{1}{L}\frac{\partial P_{old}}{\partial x}\left(0,y\right)+
\frac{1}{2L^2} \left(y -\frac{L}{2}\right)^2 \frac{\partial^2 P_{old}}{\partial x \partial y}\left(0,0\right),
\end{equation}
\begin{equation}\label{b(x)}
b(x) = \frac{1}{L}\frac{\partial P_{old}}{\partial y}\left(x,0\right)+
\frac{1}{2L^2} \left(x -\frac{L}{2}\right)^2 \frac{\partial^2 P_{old}}{\partial x \partial y}\left(0,0\right).
\end{equation}
Taking into account that 
\begin{equation}
\frac{\partial P_{old}}{\partial x}(0,y)=-\frac{\partial P_{old}}{\partial x}(L,y) \;\; , \;\; 
\frac{\partial P_{old}}{\partial y}(x,0)=-\frac{\partial P_{old}}{\partial y}(x,L),
\end{equation}
\begin{equation}
\frac{\partial^2 P_{old}}{\partial y\partial x}(0,y)=-\frac{\partial^2 P_{old}}{\partial y\partial x}(L,y) \;\; , \;\; 
\frac{\partial^2 P_{old}}{\partial x \partial y}(x,0)=-\frac{\partial^2 P_{old}}{\partial x \partial y}(x,L),
\end{equation}
it can be easily checked that $(\partial P_{corr}/\partial x)(0,y)=(\partial P_{corr}/\partial x)(L,y)$ and 
$(\partial P_{corr}/\partial y)(x,0)=(\partial P_{corr}/\partial y)(x,L)$. We notice that, for the values of parameters used in the model, the corrective terms in Eq. (\ref{Pcorr}) are much smaller than the original expression (\ref{press0}). We verified that this procedure essentially removes numerical problems related to the fulfilment of periodicity.
From now on, the expression of the equilibrium pressure will be given by $P_{corr}(x,y)$, even though we continue to indicate it by $P_0(x,y)$ to simplify the notation. A similar procedure was used in \citet{vasconez2015}. 

\begin{figure}  
\centering
\includegraphics[width=0.46\textwidth]{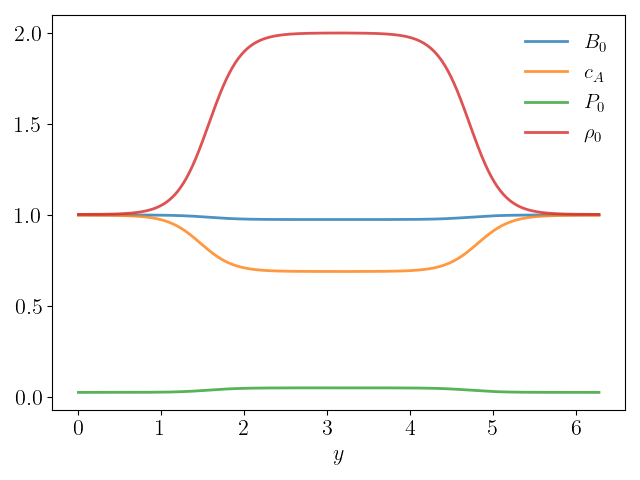}
\caption[...]
{\footnotesize{} 1D profiles of the equilibrium quantities $B_0, c_A, P_0$, and $\rho_0$ as functions of $y$, for $x=L/2$ and $z=\Lambda L/2$. }
\label{icf}
\end{figure}

The magnetic field is obtained  from Equation (\ref{equi}) as a function of $P_0$: 
\begin{equation}
\label{mag0}
B_0(x,y)=\sqrt{B_{ext}^2 + 2P_{ext} -2P_0(x,y)}.
\end{equation}
Finally, the equilibrium density is proportional to $P_0(x,y)$ according to:
\begin{equation}\label{rho0}
\rho_0(x,y)= \frac{2 P_0(x,y)}{\beta T_0}
\end{equation}
In particular, it is $P_{ext}=\beta \rho_{ext} T_0/2$ and $P_{int}=\beta \rho_{int} T_0/2$. 

To completely define the equilibrium model, we have to specify the values of parameters $\rho_{ext}$, $\rho_{int}$, $B_{ext}$, and $T_0$. We used the values: $\rho_{ext}=B_{ext}=T_0=1$, and $\rho_{int}=2$ so that the density inside the flux tube is assumed to be a factor 2 larger than the external density. For $\beta=0.05$, this gives the values $P_{ext}=0.025$ and $P_{int}=0.05$ for the plasma pressure outside and inside the flux tube, respectively.
Another relevant quantity is the Alfv\'en velocity, which is defined by $c_A(x,y)=B_0(x,y)/\sqrt{\rho_0(x,y)}$ in the code units, and varies perpedicularly to $\bm{B}_0$ 

Fig. \ref{icf} shows 1D profiles of the equilibrium quantities, $B_0(L/2,y)$, $c_A(L/2,y)$, $P_0(L/2,y)$, and $\rho_0(L/2,y)$ as functions of $y$. Those profiles are taken along a line that crosses the flux tube through the symmetry axis. The variations of density and pressure across the flux tube are clearly visible. Due to the low value of $\beta$, magnetic pressure dominates plasma pressure, and the magnetic field profile is nearly constant, though $B_0$ is slightly less intense inside the flux tube. The Alfv\'en velocity inside the flux tube is lower than outside by a $\sim 0.7$ factor. This variation is at the base of the phase-mixing phenomenon.

Finally, we notice that the unit time (in code units) corresponds to the Alfv\'en time $d_{FT}/(\pi c_{A, ext})=1$, where $d_{FT}=\pi$ is of the order of the flux tube diameter (Fig. \ref{icf}).

\subsection{Alfv\'enic torsional wave perturbation}
\label{Aws}

\begin{figure}  
\centering
\includegraphics[width=0.46\textwidth]{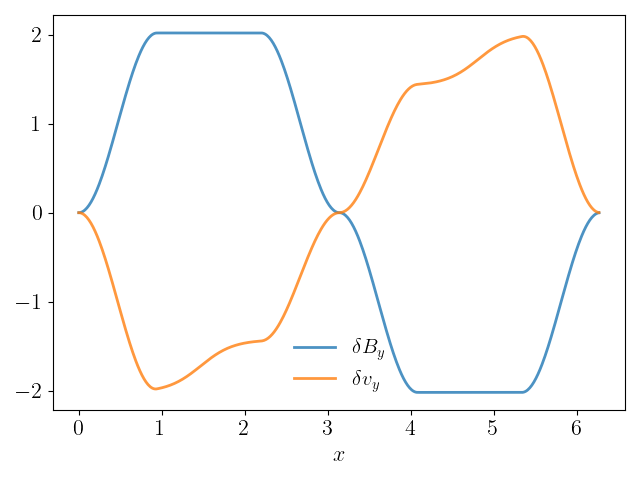}
\caption[...]
{\footnotesize{} 1D cut representing the components $\delta B_y$ and $\delta v_y$ as functions of $x$, for the single Alfv\'en wave perturbation. The section is computed along a straight line parallel to the $x$ axis, at $y=L/2$ and $z=\Lambda L/2$.}
\label{b0}
\end{figure}

The initial perturbation considered at first is a torsional Alfv\'en wave. This wave is polarized in the azimuthal direction with respect to the flux tube axis. We define the azimuthal angle $\theta=\tan^{-1}\left[ (y-y_0)/(x-x_0) \right]$ and the corresponding unit vector $\hat{\bm{\theta}}=-\sin \theta\, \hat{\bm{x}} + \cos \theta\, \hat{\bm{y}}$. The perturbation involves only the transverse components of the magnetic and velocity fields. The magnetic field perturbation reads
\begin{equation}\label{dbaw0}
\delta{\bm B}^{\text{Aw}} = A_1 f(r) \cos(k_z z) \hat{\bm{\theta}}
\end{equation}
where the superscript "Aw" indicates quantities relative to the torsional Alfv\'en wave, $A_1$ is a parameter used to fix the amplitude of the perturbation, $k_z = 2 \pi /(\Lambda L)$ represents the wavevector in the parallel $z$ direction, and $f(r)$ is a function defining the radial profile of the perturbation. In particular, we choose $f(r)$ as a \textit{tukey window} function, that has the form: 
\begin{subequations} 
\begin{numcases}{} 
f(r) = \frac{1}{2}\left[1- \cos \left( \frac{2 r}{a} \right) \right] & \text{  $0<r< a \pi/2$ },
\\[8pt]
f(r) = 1  & \text{  $a \pi/2 \leq r \leq \pi/2$ },
\end{numcases}
\end{subequations}
with $a=0.6$ and $f(\pi-r) = f(r)$ for $r< \pi/2$. This function guarantees that the perturbation smoothly goes to zero at the domain's edge and on the loop axis $r=0$. We notice that the considered torsional wave has a parallel wavelength that coincides with the parallel domain size: $\lambda_{||}=\Lambda L$ and does not depend on the azimuthal angle $\theta$, i.e., it is a $m=0$ mode. The magnetic field perturbation can be expressed in terms of Cartesian components in the form:
\begin{equation}
\label{dbaw}
\delta{\bm B}^{\text{Aw}}=\delta B_x^{\text{Aw}}\hat{{\bm x}}+\delta B_y^{\text{Aw}}\hat{{\bm y}} =
A_1 f(r)  \cos(k_z z) \left( -\sin\theta \hat{{\bm x}} + \cos\theta \hat{{\bm y}} \right)
\end{equation}
The initial perturbation is an Alfv\'en wave propagating in the positive $z$ direction. Therefore, the velocity field has the form:
\begin{equation}\label{dvaw}
    \delta \bm{v}_{x,y}^{\text{Aw}}(x,y) = -\frac{c_A(r)}{B_0(r)} \delta  \bm{B}_{x,y}^{\text{Aw}}(x,y) = -\frac{ \delta  \bm{B}_{x,y}^{\text{Aw}}(x,y)}{\sqrt{\rho_0(r)}},
\end{equation}
Fig. \ref{b0} shows 1D profiles of the $\delta{\bm B}^{\text{Aw}}_y(x, L/2)$ and $\delta{\bm v}^{\text{Aw}}_y(x, L/2)$ components taken along a line that crosses the flux tube axis parallel to the $x$ axis (in Fig. \ref{b0} we set $A_1=1$). Moreover, a 2D plot of $\delta{\bm B}^{\text{Aw}}_y$ in a plane perpendicular to $\bm{B}_0$ is represented in the left panel of Fig. \ref{alfvenevo}.
Both magnetic field and velocity perturbations are mainly localized across the shear region, but they are also present in the remaining part of the domain, compatible with the Cartesian geometry of $D$.
The value of the parameter $A_1$ is determined by imposing that the RMS value of the initial magnetic field perturbation is unitary:
$\delta B_{\perp, \text{rms}} \equiv \sqrt{ \langle (B_x - \langle B_x \rangle)^2 + (B_y - \langle B_y \rangle)^2 \rangle}=1$,
where $\langle ... \rangle$ indicates a volume average. The actual RMS in the production runs is obtained by multiplying the perturbed field by a specific value depending on the run.
The perturbation does not affect the remaining physical quantities: $\delta \rho=\delta v_z=\delta B_z=\delta T=0$. 

Eqs. (\ref{dbaw}) and (\ref{dvaw}) correspond to a single Alfv\'en wave propagating along the background magnetic field. However, we will also consider the case of a standing Alfv\'en wave. This can be obtained starting from an initial condition where the magnetic field perturbation has the same form (\ref{dbaw}), but the velocity perturbation is initially set to zero. This situation corresponds to two counter-propagating Alfv\'en waves initially having opposite velocity fields that cancel each other. While the propagating wave perturbation (denoted by "Pw", propagating wave) involves both magnetic and kinetic energy (that are equal for an Alfv\'en wave), the standing wave perturbation (denoted by "Sw", standing waves) initially involves only magnetic energy. On the other hand, for the sake of comparison, it is appropriate that the two kinds of perturbations have the same amount of fluctuating energy. This condition is satisfied by choosing the parameter $A_1$ such that $A_1^{\textbf{Sw}} = \sqrt{2} \, A_1^{\textbf{Pw}}$.

\subsection{Turbulent perturbation}
\label{turb}
Another form of initial perturbation considered here corresponds to a turbulent perturbation. The corresponding magnetic field $\delta \bm{B}^\text{turb}$ and velocity $\delta \bm{v}^\text{turb}$ perturbations are written as a superposition of transverse Fourier modes, each polarized in the $xy$--plane and characterized by a wavevector ${\bm k}=(k_x,k_y,k_z)$. Both $\delta \bm{B}^\text{turb}$ and $\delta \bm{v}^\text{turb}$ are divergence-free, the latter condition to ensure that the initial perturbation is non-compressive, as for Alfv\'enic perturbations. 
We describe a procedure to construct $\delta \bm{B}^\text{turb}$ and $\delta \bm{v}^\text{turb}$ with the aforementioned characteristics. 

To comply with the divergence-free condition, the turbulent magnetic perturbation is expressed as $\delta \bm{B}^\text{turb}=\nabla \times \bm{A}$, with $\bm{A}$ the corresponding vector potential. 
Since we are interested in fluctuations polarized in the direction perpendicular to ${\bm B}_0$, we chose a vector potential having the form ${\bm A}={\bm A}(x,y,z)=A_x(z) \hat{\bm{x}} + A_y(z) \hat{\bm{y}} + A_z(x,y,z)) \hat{\bm{z}}$, in which the $A_x$ and $A_y$ components account for modes with $\bm{k}$ parallel to $\bm{B}_0$ and the $A_z$ component accounts for oblique and perpendicular modes.
Periodicity allows us to write the vector potential in terms of a Fourier series as
$A_x = \sum_{k_z}\hat{A}_x({k_z}) e^{i{k_z}{z}} $, 
$A_y = \sum_{k_z}\hat{A}_y({k_z}) e^{i{k_z}{z}} $,
$A_z = \sum_{k_x,k_y,k_z}\hat{A}_z(k_x,k_y) e^{i{\bm k}\cdot{\bm x}} $, where the wavevector is $\bm{k}=[(2\pi/L) n_x , (2\pi /L)n_y , (2\pi /\Lambda L)n_z ]$, $n_x$, $n_y$ and $n_z$ are integers, and $(n_x,n_y,n_z)$ is the wavenumber. The corresponding magnetic field has the form:
$$\delta B_x^{\text{turb}} = \sum_{k_x,k_y,k_z}ik_y\hat{A}_z({\bm k}) e^{i{\bm k}\cdot{\bm x}} - 
        \sum_{k_z}ik_z\hat{A}_y({k_z}) e^{i{k_z}{z}} $$
$$\delta B_y^{\text{turb}} = \sum_{k_z}ik_z\hat{A}_x({k_z}) e^{i{k_z}{z}} - 
        \sum_{k_x,k_y,k_z}ik_x\hat{A}_z({\bm k}) e^{i{\bm k}\cdot{\bf x}} $$
$$\delta B_z^{\text{turb}} = 0,$$
The Fourier coefficients $\hat{A}_x,\hat{A}_y,\hat{A}_z$ are chosen such that the spectrum of the initial fluctuations is isotropic in the wavenumber space. In particular, for $\sqrt{n_x^2+n_y^2+n_z^2}\le n_{max}$, it is $|\hat{A}_x(k_z)| = |\hat{A}_y(k_z)| \propto |k_z|^{-a}$ and $|\hat{A}_z(k_x,k_y)| \propto k_\perp^{-a}$, with $k_\perp=\sqrt{k_x^2+k_y^2}$. Otherwise, for $\sqrt{n_x^2+n_y^2+n_z^2}> n_{max}$ it is $\hat{A}_x(k_z)=\hat{A}_y(k_z)=\hat{A}_z(k_x,k_y)=0$.
We used the values $a=1.5$ and $n_{max}=3$, giving an initial perturbation localized at the largest spatial scales. The phases of complex Fourier coefficients $\hat{A}_x$, $\hat{A}_y$, $\hat{A}_z$ are randomly chosen.

The same procedure is applied to obtain the velocity fluctuation $\delta \bm{v}^\text{turb}$. However, the phases of the velocity Fourier coefficients are chosen differently with respect to the magnetic field. In particular, phases are chosen such as the cross helicity $H_c=\int\delta{\bm v}^\text{turb} \cdot \delta{\bm B}^\text{turb} \;dV$ of the initial perturbation is small ($\sim 1\%$) with respect to the fluctuating magnetic energy $E_M=\int\left(\delta{\bm B}^\text{turb}\right)^2/2 \;dV$. This condition is necessary in order to trigger a nonlinear cascade, at least in the case of a uniform background.
Finally, as for the Alfv\'enic torsional wave, both magnetic and velocity fluctuation are normalized to their respective RMS values.

The pattern of the above-described turbulent perturbation is represented in the upper panel of Fig. \ref{turb_c}, where a 3D visualization of the $B_y$ component at the initial time $t=0$ is drawn in the case $\Lambda=4$. 

\section{Numerical results}
\label{sec3}
\begin{figure}[t]
\hspace{-0.3cm}
\includegraphics[width=0.5\textwidth]{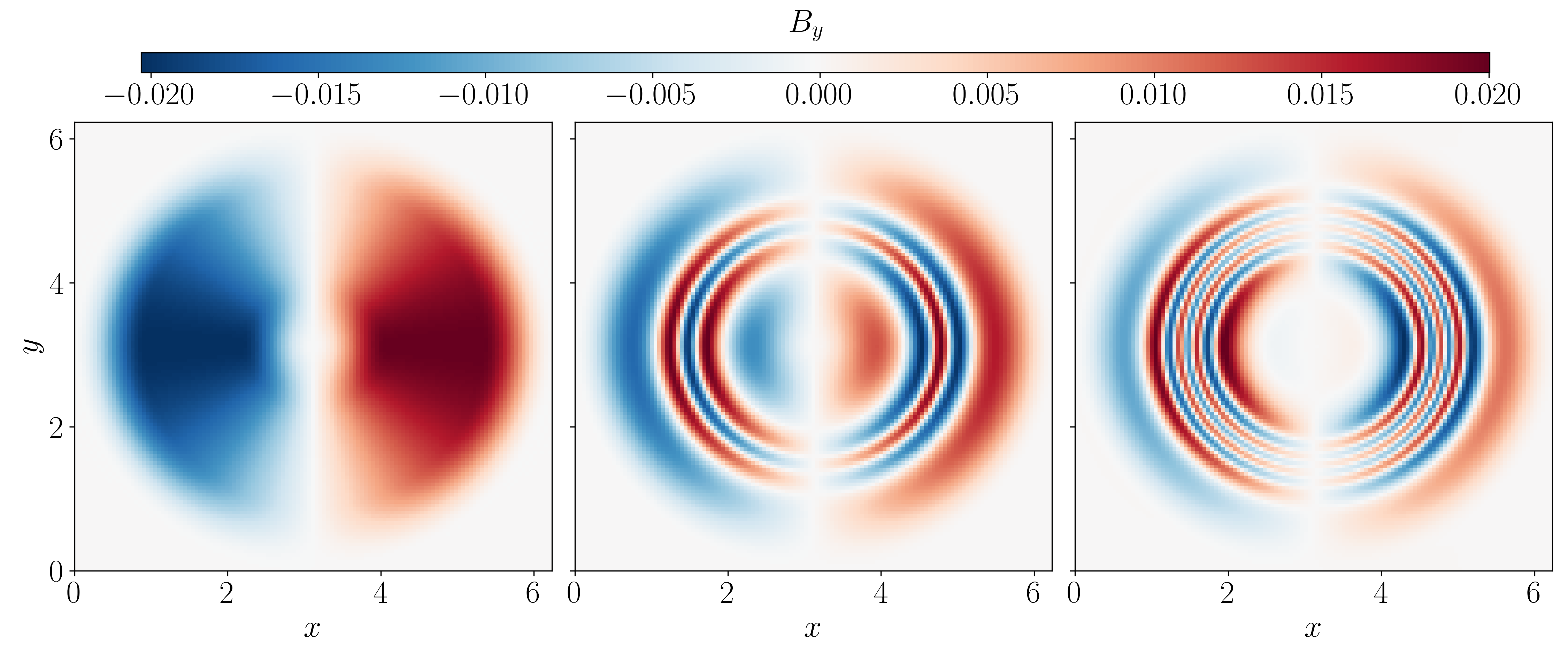}
\caption[...]
{\footnotesize{} 2D perpendicular section of the $B_y$ component for the Alfv\'en waves evolution at times $t=0$ (left), $t=150$ (middle), and $t=300$ (right). Here we used a perturbation amplitude $A=10^{-2}$, a hyper-dissipation coefficient $\eta=10^{-8}$, and an aspect ratio $\Lambda = 4$ (Run 15). }
\label{alfvenevo}
\end{figure}

\begin{table}
\caption{Table of simulations. From left to right are reported: a number identifying the Run, the initial perturbation type, the amplitude A, the number of mesh points in the perpendicular and parallel directions, the aspect ratio $\Lambda$, the hyper-dissipative coefficient $\eta$, and the parameter $\alpha$, when appropriate. For all the simulations, we used a plasma-$\beta = 5 \times 10^{-2}$. In the second column, "IEq" and "HEq" indicate the inhomogeneous or homogeneous equilibrium, respectively, "Pw" and "Sw" indicate the propagating and the standing Alfvén wave, respectively, and "Turb" the turbulent perturbation. All runs are 3D.} 
\setlength{\tabcolsep}{4.5pt}
\label{table1} 
\centering 
\begin{tabular}{c c c c c c c c} 
\hline\hline 
Run & IC type & A & $N_\perp$ & $N_\parallel$ & $\Lambda$ & $\eta$ & $\alpha$ \\ 
\hline 
1 & IEq+Pw & 0.02 & 64 & 64  & 1 & $5 \times 10^{-6}$ & - \\ 
2 & IEq+Pw & 0.05 & 64 & 64  & 1 & $5 \times 10^{-6}$ & - \\ 
3 & IEq+Pw & 0.1 & 64 & 64  & 1 & $5 \times 10^{-6}$ & - \\ 
4 & IEq+Pw & 0.2 & 64 & 64  & 1 & $5 \times 10^{-6}$& - \\ 
5 & IEq+Pw & 0.02 & 64 & 64  & 2 & $5 \times 10^{-6}$& - \\ 
6 & IEq+Pw & 0.02 & 64 & 64  & 3 & $5 \times 10^{-6}$& - \\ 
7 & IEq+Pw & 0.02 & 64 & 64  & 4 & $5 \times 10^{-6}$& - \\ 
8 & IEq+Pw & 0.02 & 64 & 64  & 5 & $5 \times 10^{-6}$& - \\ 
9 & IEq+Pw & 0.02 & 64 & 64  & 6 & $5 \times 10^{-6}$& - \\ 
10 & IEq+Pw & 0.02 & 64 & 64  & 7 & $5 \times 10^{-6}$& - \\ 
11 & IEq+Pw & 0.02 & 64 & 64  & 8 & $5 \times 10^{-6}$& - \\ 
12 & IEq+Pw & 0.02 & 64 & 64  & 9 & $5 \times 10^{-6}$& - \\ 
13 & IEq+Pw & 0.02 & 64 & 64  & 10 & $5 \times 10^{-6}$& - \\ 
14 & HEq+Turb & 0.01 & 512 & 16  & 4 & $10^{-8}$& - \\ 
15 & IEq+Pw & 0.01 & 512 & 16  & 4 & $10^{-8}$& 0.0 \\ 
16 & IEq+Pw+Turb & 0.01 & 512 & 16  & 4 & $10^{-8}$& 0.23 \\ 
17 & IEq+Pw+Turb & 0.01 & 512 & 16  & 4 & $10^{-8}$& 0.5 \\ 
18 & IEq+Pw+Turb & 0.01 & 512 & 16  & 4 & $10^{-8}$& 0.66 \\ 
19 & IEq+Pw+Turb & 0.01 & 512 & 16  & 4 & $10^{-8}$& 0.83 \\ 
20 & IEq+Pw+Turb & 0.01 & 512 & 16  & 4 & $10^{-8}$& 1.0 \\ 
21 & IEq+Sw & 0.01 & 512 & 16  & 4 & $10^{-8}$& 0.0 \\ 
22 & IEq+Sw+Turb & 0.01 & 512 & 16  & 4 & $10^{-8}$& 0.23 \\ 
23 & IEq+Sw+Turb & 0.01 & 512 & 16  & 4 & $10^{-8}$& 0.5 \\ 
24 & IEq+Sw+Turb & 0.01 & 512 & 16  & 4 & $10^{-8}$& 0.66 \\ 
25 & IEq+Sw+Turb & 0.01 & 512 & 16  & 4 & $10^{-8}$& 0.83 \\ 
26 & IEq+Sw+Turb & 0.01 & 512 & 16  & 4 & $10^{-8}$& 1.0 \\ 
\hline 
\end{tabular}
\end{table}

In this section, we present results from numerical simulations. 
In order to quantify the evolution of the dissipation rate in our model, we monitored the evolution of the quadratic quantity $W=\langle J^2 \rangle + \langle \omega^2 \rangle$ in time, where $\bm{J}=\nabla \times \bm{B}$ and $\bm{\omega}=\nabla \times \bm{v}$ are the rms values of the current density and vorticity, respectively.
For each run, we define the dissipation time $t_d$ as the time that corresponds to the maximum dissipation rate, namely $W_{\text{max}} = W(t=t_d)$. 
To properly quantify the value of $t_d$, we carried out all simulations up to a time $t_\text{final}$ when $W$ becomes definitely smaller than $W_{\text{max}}$.
Each run has been performed using the same value $\eta$ for the hyper-dissipative coefficients: $\eta=\nu_4=\eta_4=\chi_4$, even though $\eta$ can be different in different runs. 
In the following, we will refer to the fields in the $xy-$plane with the subscript $``\perp"$, while $``\parallel"$ represents the parallel component to the $z-$axis.
We also remind that the Alfv\'en time is unitary in our normalisation units.
All the simulations performed are summarized in table \ref{table1} where we report the values of various parameters.

\begin{figure}[t]
\centering
\includegraphics[scale=0.52]{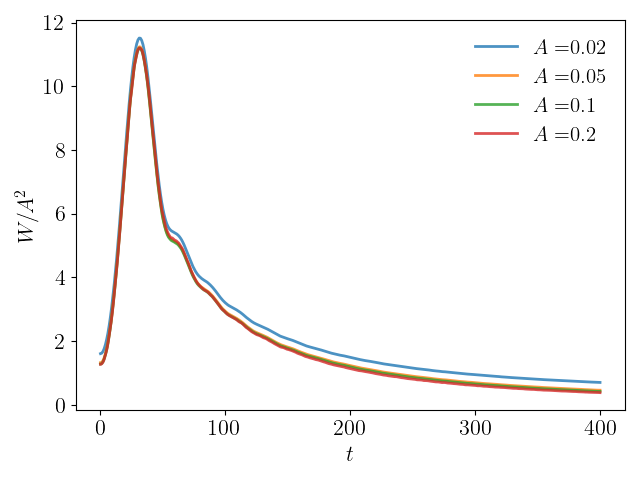}
\caption[...]
{\footnotesize{} Time history of $W/A^2$ for the propagating Alfv\'en wave, with different values of amplitude A, namely for Runs 1 -- 4.}
\label{diff_A.png}
\end{figure}

\subsection{Torsional Alfv\'en wave}
\label{sec3.1}

We performed a series of runs to characterize the time evolution of the torsional Alfv\'en wave propagating in the inhomogeneous equilibrium, described in Sect.s \ref{sec2.3} and \ref{Aws}, respectively.  The main purpose is to verify the fulfilment of scaling laws characterizing phase mixing by the numerical model. 
We initialize the fields as:  
\begin{eqnarray}
    B_\parallel &=& B_0(x,y),
    \label{ic1}
    \\
    \label{bperp}
    \bm{B}_\perp &=&  A \, \delta \bm{B}_\perp^{\text{Aw}}(x,y,z),
    \\
    v_\parallel &=& 0,
    \\
    \bm{v}_\perp &=& A \, \delta \bm{v}_\perp^{\text{Aw}}(x,y,z),
    \label{ic4}
    \\
    \rho &=& \rho_0(x,y).
    \label{ic5}
\end{eqnarray}
Due to the normalization of $\delta \bm{B}_\perp^{\text{Aw}}$ described in Section \ref{Aws}, the parameter $A$ represents the rms amplitude of 
the fluctuating magnetic field.

In Fig. \ref{alfvenevo} we report a 2D perpendicular section at $z=\Lambda L /2$ of the $B_y$ component for the high-resolution Run 15, where only the torsional wave is initially present (Table \ref{table1}), at time $t=0$ (left), $t=150$ (middle), and $t=300$ (right).
The last time approximately corresponds to the dissipation time $t_d$ for this test. 
As the simulation proceeds, phase mixing occurs, generating increasingly smaller length scales in the radial direction (across the background magnetic field). This gradually increases $W$, until dissipative effects due to hyper-dissipation terms come into play, eroding the fluctuation and reducing $W$. This happens all across the boundary of the flux tube, where the inhomogeneity is localized.

In order to quantitatively characterize the phenomenon of phase mixing, we can derive an estimation for the dissipative time $t_d$ in numerical runs. At variance with Eq. (\ref{tdPM0}) where $t_d$ has been expressed in terms of the dissipative scale $\ell_d$, in the numerical model, such a scale is not fixed a priori. Instead, dissipation is determined by the hyper-dissipative terms included in the model. In particular, we can define the characteristic time $t_\eta$ associated with hyper-dissipation by balancing the term $\partial \bm{A}/\partial t$ with $\eta \nabla^4 \bm{A}$ in Eq. (\ref{MHD3}) (or, equivalently, balancing $\partial \bm{v}/\partial t$ with $\eta \nabla^4 \bm{v}$ in Eq. (\ref{MHD2})). This gives
\begin{equation}
    t_\eta(k_\perp) \sim \frac{1}{\eta k_\perp^4}.
    \label{t_eta}
\end{equation}
The dynamical time associated with phase mixing is given by the expression (\ref{tPM0}) where, in our case, $k_{||0}=2\pi/(\Lambda L)=\Lambda^{-1}$. This gives:
\begin{equation}\label{tPM1}
t_{PM}(k_\perp) \sim k_\perp \Lambda \left( \frac{d c_A}{dr} \right)^{-1}
\end{equation}
Dissipation becomes effective at a wavevector $k_{\perp d}$ where $t_{PM}(k_\perp)$ becomes of the order of $t_\eta(k_\perp)$. Therefore, matching Eq.s (\ref{t_eta}) and (\ref{tPM1}) we derive:
\begin{equation}\label{kperpd}
k_{\perp d} \sim \frac{1}{(\eta \Lambda)^{1/5}} \left( \frac{dc_A}{dr}\right)^{1/5}
\end{equation}
Inserting this expression into Eq. (\ref{tPM1}) we derive an estimation for the dissipative time for the torsional Alfv\'en wave case:
\begin{equation}
    t_d \sim \left( \frac{dc_A}{dr}\right)^{-4/5} \eta^{-1/5}  \Lambda^{4/5},
    \label{td}
\end{equation}
Moreover, estimating the current density as $J\sim k_\perp B_\perp$, and assuming that phase mixing does not substantially modify the initial wave amplitude $A$ as long as $k_\perp \lesssim k_{\perp d}$, we can derive an estimation for the maximum dissipation rate:
\begin{equation}\label{J2max}
     J^2_{\text{max}} \sim A^2 \left( \frac{1}{\eta \Lambda} \frac{dc_A}{dr}   \right)^{2/5}.   
\end{equation}

Eqs. (\ref{td}) and (\ref{J2max}) can be compared with the results of numerical runs in the case of the torsional Alfv\'en wave. In particular, the dependence of the dissipation on the wave amplitude has been studied by performing some numerical tests (Run 1 to Run 4) where different values of $A$ have been used, varying $A$ over one order of magnitude: $A=0.02, 0.05, 0.1, 0.2$. Since an high resolution is not required for these runs, we used a grid with $N_\perp=N_{||}=64$ mesh points for each direction. The domain $D$ is a cube  ($\Lambda=1$) of size $L = 2\pi$. Due to the low number of gridpoints, the hyper-dissipative coefficients have the relatively high value $\eta=5 \times 10^{-6}$. 
In Fig. \ref{diff_A.png}, we report the time history of $W/A^2$ for those runs (Run 1-4). It appears that the four curves are almost superposed, indicating that the dissipation rate $W$ is proportional to $A^2$, in accordance with Eq. (\ref{J2max}). Moreover, the dissipative time $t_d$, corresponding to the time of maximum $W$, is independent of the wave amplitude $A$, as predicted by Eq. (\ref{td}). 

In Fig. \ref{diff_A.png} the normalized profile $W(t)/A^2$ corresponding to the lowest amplitude $A=0.02$ is slightly higher than the others. This small discrepancy can be attributed to the power developed by the dissipation of the current associated with the equilibrium magnetic field $\bm{B}_0(x,y)$. Of course,  for low values of the wave amplitude $A$ this effect is becomes more visible in comparison with wave dissipation rate. However, even for low wave amplitudes the dissipation of the equilibrium current remains sufficiently small to be ignored. We also observe that the dissipation of the equilibrium current is completely negligible in the high-resolution runs (Run 14--26) described in Sec.s \ref{append}--\ref{sec3.4}, where a much lower value for the dissipative coefficient $\eta$ has been used. 

\begin{figure}
\centering
\includegraphics[scale=0.5]{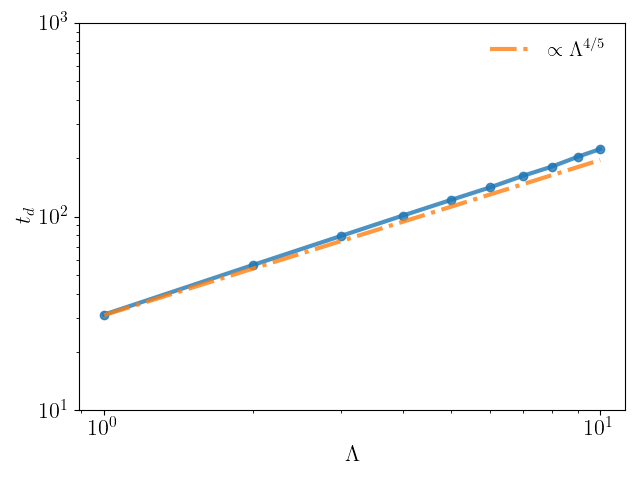}
\caption[...]
{\footnotesize{} Dissipation time $t_d$ for different aspect ratio, namely $\Lambda \in [1,10]$. The dash-dot line represents a power law with slope $4/5$, according to our estimation in equation (\ref{td}). These simulations are reported in table \ref{table1} as Run 1 and Run 5--13.}
\label{lpar}
\end{figure}
Another set of low-resolution runs has been performed, where the aspect ratio $\Lambda$ of the spatial domain has been increased from $\Lambda=1$ up to $\Lambda=10$. For all those runs, we set an amplitude $A=0.02$, the coefficient $\eta=5 \times 10^{-6}$, and we used $N=64$ mesh points for each direction. These simulations are reported in table \ref{table1} as Run 1 and Run 5 -- 13. For each run, we calculated the dissipative time $t_d$ as the time of maximum $W$.
In Fig. \ref{lpar}, we report the dissipative time as a function of the aspect ratio $\Lambda$ (dotted line). It appears that the dissipative time follows the law $t_d \propto \Lambda^{4/5}$ (dotted-dashed line) reasonably well, in accordance with what is predicted by Eq.~(\ref{td}).  

A closer examination of Fig. \ref{lpar} reveals that the dissipative time increases with $\Lambda$ slightly faster than the prediction $t_d \propto \Lambda^{4/5}$. Such a small discrepancy can be explained by taking into account that Eq. (\ref{td}) has been derived assuming a radial profile of the Alfv\'en velocity that remains unchanged during the wave evolution. Instead, dissipation also affects the equilibrium magnetic field slightly reducing the gradient of the Alfv\'en velocity in time, as we explicitly verified. According to Eq. (\ref{td}), a decrease of $d c_A/dr$ corresponds to an increase of $t_d$. This effect is more pronounced when $t_d$ is larger (i.e., for larger $\Lambda)$ since, in this case, dissipation has more time to smooth the profile of $c_A(r)$. This corresponds to a deviation from the scaling law $t_d \propto \Lambda^{4/5}$ that is larger for larger $\Lambda$.

Summarizing, the above results show that, in the configuration considered in  our model, the time evolution of the torsional Alfv\'en wave is fully dominated by phase mixing, and the corresponding properties are well reproduced by the numerical model. On the contrary, no relevant nonlinear effects have been observed on transverse fluctuations characterizing this wave, even for relatively large values of the wave amplitude ($A=0.2$).

\subsection{Turbulence in homogeneous background}
\label{append}

\begin{figure}
\includegraphics[scale=0.25]{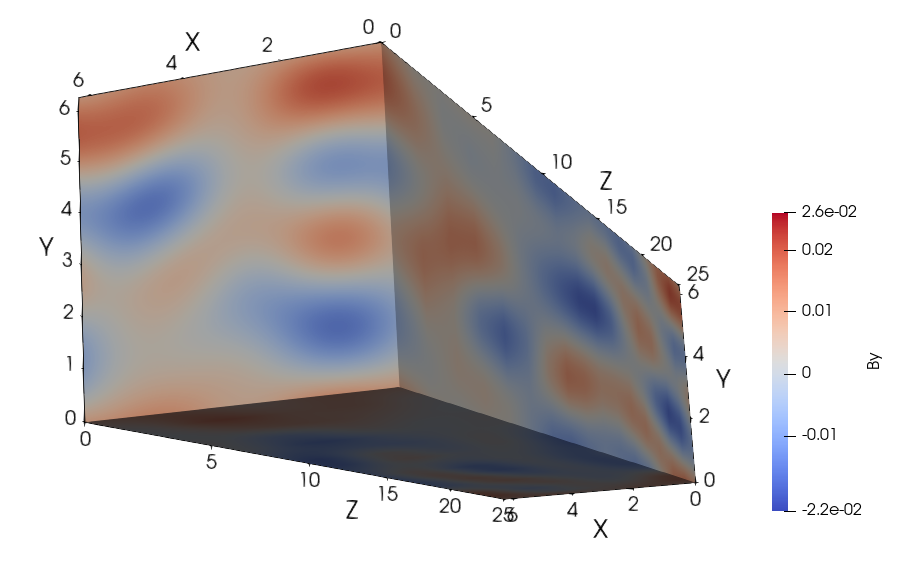}
\includegraphics[scale=0.25]{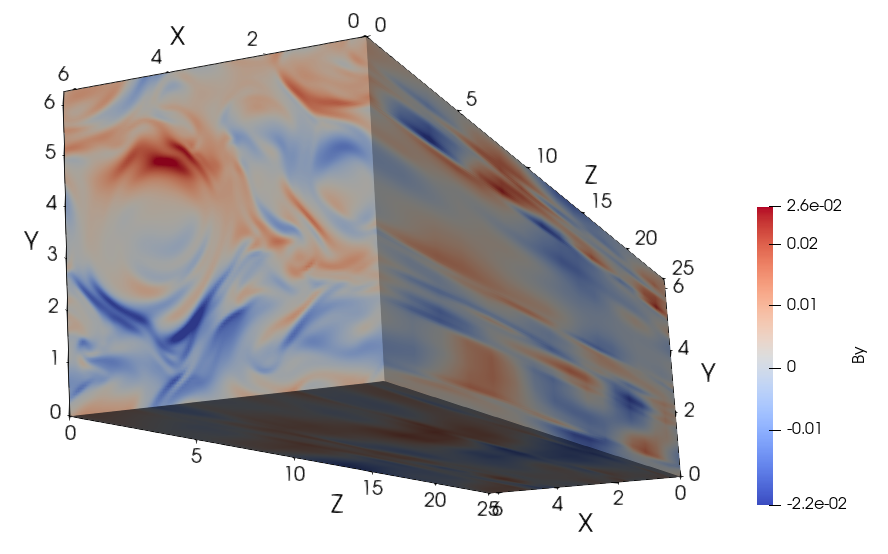}
\caption[...]
{\footnotesize{} 3D visualization of the $B_y$ component at times $t=0$ (top) and $t=t_d$ (bottom), for Run 14. }
\label{turb_c}
\end{figure}

In this section, we describe the time evolution of the turbulent perturbation, defined in Sec. \ref{turb}, in a homogeneous background (Run 14). Though the main focus of this paper is on inhomogeneous structures, here, the main purpose is to fix a reference case. 
In Run 14, we used a low value of the hyper-dissipative coefficient ($\eta=10^{-8}$) and a high spatial resolution in the perpendicular direction, trying to obtain a spectrum with a wide range of scales compatible with numerical limitations. The results of Run 14 will be used in Sect. \ref{sec3.3} for a comparison with the behavior of waves and turbulence in the inhomogeneous equilibrium (Runs 15-26). Therefore, in Run 14 we use the same values for the parameters $A$, $\Lambda$, and $\eta$ as in Runs 15--26. In particular, we set a low value for the fluctuation amplitude ($\delta B/B_0\simeq A=0.01$). Such a choice will allow us to satisfy the condition (\ref{cond}) for Runs 15--26 of Section \ref{sec3.3}. Of course, lower values for $A$ would better fulfil condition (\ref{cond}), but this would be exceedingly costly due to long simulation times. We also observe that a situation characterized by small fluctuation amplitude $\delta B/B_0$ and low dissipation can be considered to be more representative of the corona.

In Run 14, physical quantities have been initialized in the following way: $B_{||}=B_0=1$, $\rho=\rho_0\simeq 1.21$, which is the average value of $\rho_0$ in the inhmogeneous cases, and $c_A\simeq 0.91$. Moreover, $\bm{B}_\perp=A\, \delta \bm{B}_\perp^{\text{turb}}$, $\bm{v}_\perp=A\, \delta \bm{v}_\perp^{\text{turb}}$ and $v_{||}=0$, where the turbulent perturbations $\delta \bm{B}_\perp^{\text{turb}}$ and $\delta \bm{v}_\perp^{\text{turb}}$ have been specified in Sect. \ref{turb}. A 3D visualization of $B_y$ corresponding to the above initial condition is plotted in Fig. \ref{turb_c}, upper panel. It can be observed that the typical transverse scale associated with fluctuations is $\ell_{\perp 0} \sim 1$ (in code units).

\begin{figure}
	\centering
	\includegraphics[scale=0.5]{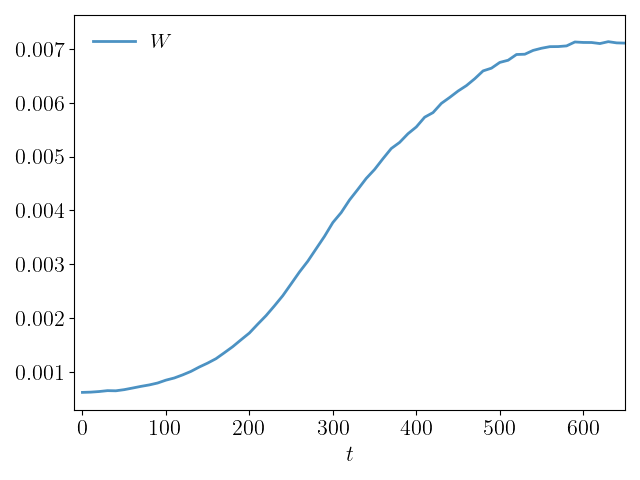}
	\caption[...]
	{\footnotesize{} Time history of $W$ for the turbulence in a homogeneous background with low amplitude $A$ and low hyper-dissipative coefficient $\eta$ (Run 14).}.
	\label{j2turb}
\end{figure}

In Fig. \ref{j2turb}, the quantity $W$ is plotted as a function of time for Run 14. The value $t_d\simeq 600$ can be derived for the dissipative time in this case. Using such a value in Eq. \ref{tNLMHD}, as well as $c_{A0}=1$, $\ell_{\perp 0} \sim 1$ and $\delta B(\ell_{\perp 0})/B_0=0.01$, we obtain $\Gamma \simeq 6$. This value will be used in Sect. \ref{sec3.3} to identify the regime of turbulence in an inhomogeneous background with respect to the condition (\ref{cond}). 
In Fig. \ref{turb_c}, lower panel, the $B_y$ component is represented at the time $t=t_d$. Comparing with the image at the initial time (upper panel), the presence of structures at small scales is clearly visible at $t=t_d$. Moreover, such structures are dominated by perpendicular wavevectors, as is expected for turbulence with a strong background magnetic field.

\subsection{Phase mixing and turbulence in an inhomogeneous background}
\label{sec3.3}

In this section, we aim to investigate the synergy between phase mixing and nonlinear effects in the generation of small scales within an inhomogeneous equilibrium, like the flux tube we considered. As previously seen, the evolution of the torsional Alfv\'en wave can be completely described in terms of phase mixing. However, a fluctuation exactly corresponding to an Alfv\'en torsional wave is unlikely to be present in a loop. More realistically, one can think of a torsional wave that is more or less distorted. Such a distortion can be represented as a superposed turbulent component on the wave. In this case, important points are investigating whether phase mixing is still effective for a distorted torsional wave and how dissipation depends on the relative amplitude of the turbulent component with respect to the wave. Moreover, considering the turbulent component itself, the inhomogeneity of the background structure acting on it could produce small scales through a mechanism similar to phase mixing. In this case,  both phase mixing and nonlinear cascade should contribute to the generation of small scales. This combined mechanism should work even in a case where the amplitude of the turbulent component is comparable with or larger than the torsional wave.

The above speculations have been investigated by considering a perturbation given by the superposition of a turbulent fluctuation on a torsional Alfv\'en wave and making the whole perturbation to evolve in the inhomogeneous equilibrium. Therefore, the initial condition is given by:
\begin{eqnarray}
    B_\parallel &=& B_0(x,y),
    \label{ic1}
    \\
    \bm{B}_\perp &=& A\, \delta \bm{B}_\perp^{\text{WT}}(x,y,z) \nonumber \\
   &=& A \, \left[(1-\alpha)\, \delta \bm{B}_\perp^{\text{Aw}}(x,y,z) + \alpha \delta \bm{B}_\perp^{\text{turb}}(x,y,z) \right], \label{BperpAwTurb}
    \\
    v_\parallel &=& 0,
    \\
    \bm{v}_\perp &=& A\, \delta \bm{v}_\perp^{\text{WT}}(x,y,z) \nonumber \\ 
    &=& A \, \left[(1-\alpha)\, \delta \bm{v}_\perp^{\text{Aw}}(x,y,z) + \alpha \delta \bm{v}_\perp^{\text{turb}}(x,y,z) \right],  \label{vperpAwTurb}
    \label{ic4}
    \\
    \rho &=& \rho_0(x,y),
    \label{ic5}
\end{eqnarray}
where $\alpha \in [0,1]$ is a free parameter that allows us to tune the amount of turbulence superposed on the torsional wave. Here, "WT" stands for "wave+turbulence".
For each value of $\alpha$, magnetic field $\delta \bm{B}_\perp^{\text{WT}}$ and velocity $\delta \bm{v}_\perp^{\text{WT}}$ perturbations are normalized to the rms value of $\delta \bm{B}_\perp^{\text{WT}}$. Therefore, the quantity $A$ gives the amplitude of the whole perturbation. The case $\alpha=0$ corresponds to a purely torsional Alfv\'en wave while increasing $\alpha$ the turbulence level increases. 
We performed direct numerical tests by varying $\alpha$ in two cases: (i) propagating Alfv\'en wave plus turbulence and (ii) standing Alfv\'en wave plus turbulence. The forms used for propagating and standing Alfv\'en waves have been specified in Sect. \ref{Aws}.
For all the simulations described in the Sect.s \ref{sec3.3.1} and \ref{sec3.3.2} we used a low perturbation amplitude $A=0.01$, an aspect ratio $\Lambda = 4$, a low hyper-dissipative coefficient $\eta = 10^{-8}$, and $N_{\perp} = 512$, $N_\parallel = 16$ mesh points. As already remarked, these values are the same as those used in the low dissipative, low-amplitude Run 14, where we considered only a turbulent perturbation in a homogeneous background. 

An important point is checking whether the regime of parameters we are considering (for the turbulence-dominated case $\alpha \sim 1$) is consistent with the inequality (\ref{cond}). As discussed in Sect. \ref{sec2.1}, if the inequality (\ref{cond}) is satisfied by the initial condition, then phase mixing dominates during the initial stage and nonlinear cascade during a later stage. In the present case we estimate terms in the inequality (\ref{cond}) by using the following values: $dc_A/dr \simeq 0.3$ (Fig. \ref{icf}); $c_A \simeq 1$; $L_{||}=\Lambda L=8\pi$; $\ell_{\perp 0}\sim 1$ and $\Gamma \simeq 6$ (Sect. \ref{append}). This gives the value $\simeq 0.07$ for the R.H.S of (\ref{cond}), while the L.H.S. is $\delta B/B_0 = A = 0.01$. Therefore, the choice of parameters corresponds to a regime identified by the condition (\ref{cond}), when $\alpha \sim 1$. The same regime holds even more for lower values of $\alpha$, since in cases when the torsional wave has a larger weight with respect to the turbulent component, the nonlinear cascade should play a less relevant role with respect to phase mixing.

\subsubsection{Propagating Alfv\'en wave plus turbulence}
\label{sec3.3.1}

\begin{figure}[t]
\includegraphics[scale=0.35]{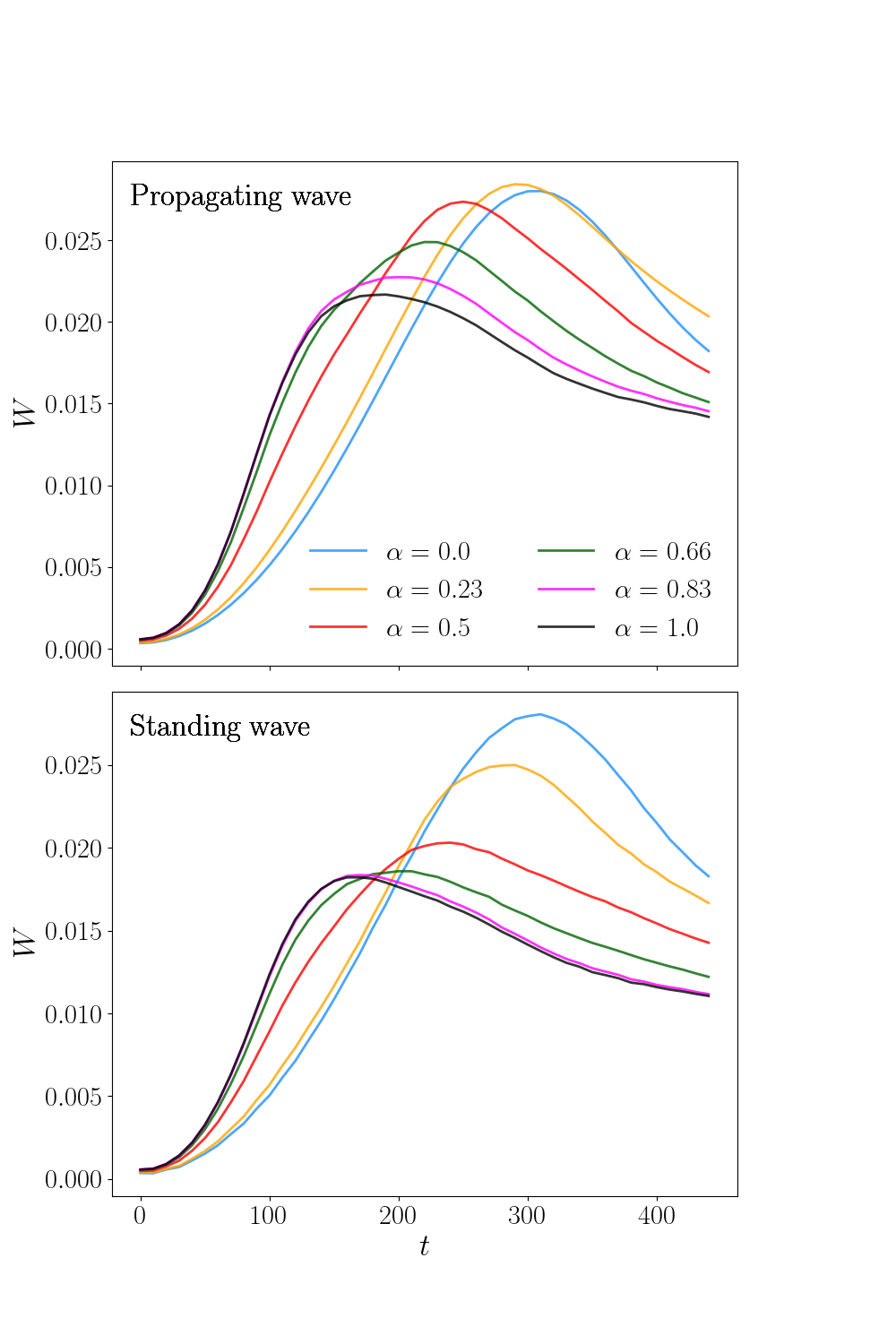}
\caption[...]
{\footnotesize{} Time history of $W$ for the case of a single Alfv\'en wave (top panel for Runs 15--20) and two Alfv\'en waves (bottom panel for Runs 21--26), and different values of $\alpha$. One can notice that the dissipative time decreases for increasing $\alpha$.}
\label{1w2w}
\end{figure}
\begin{figure}
\centering
\includegraphics[scale=0.25]{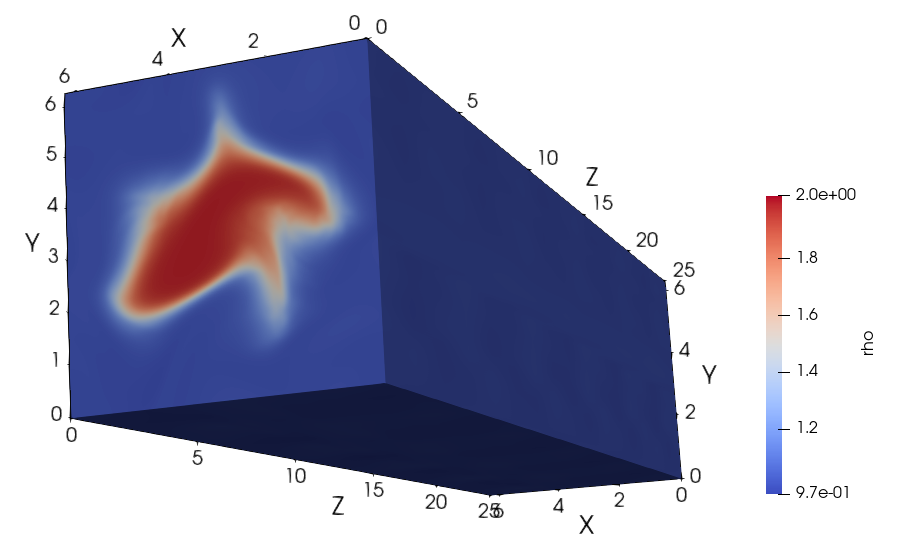}
\includegraphics[scale=0.25]{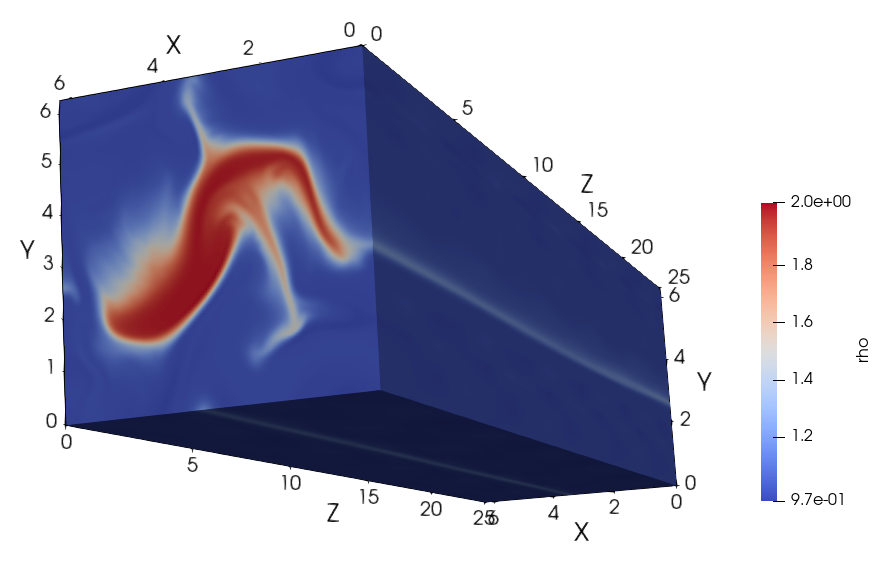}
\includegraphics[scale=0.25]{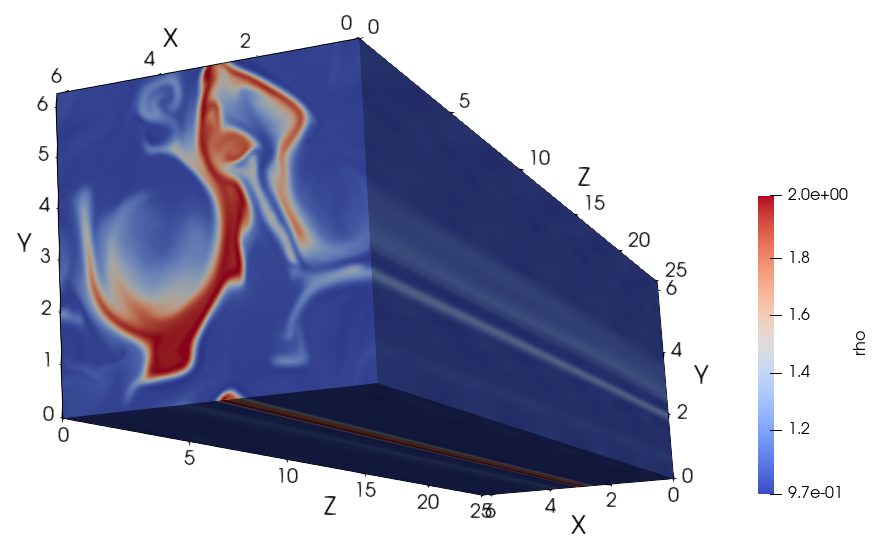}
\caption[...]
{\footnotesize{} 3D visualization for the density field at time $t = 200$ (top), $t = 450$ (center), and $t = 700$ (bottom), for a case of a turbulent perturbation propagating in the inhomogeneous equilibrium ($\alpha=1$, Run 20).}
\label{alpha2}
\end{figure}

The case of a single propagating torsional Alfv\'en wave plus turbulence has been considered in Runs 15--20, where the value of the parameter $\alpha$ has been varied between $\alpha=0$ and $\alpha=1$, thus changing the relative weight of the two components.

The time history of the dissipation rate $W$ for these cases is reported in the top panel of figure \ref{1w2w}. Each curve corresponds to a different value of $\alpha$. 
It can be observed that the dissipative time $t_d$ decreases with increasing $\alpha$. In particular, $t_d$ is reduced by a factor $\simeq 2$ going from $\alpha=0$ (torsional wave only) to $\alpha=1$ (turbulence only). As expected, the presence of a turbulent component superposed on the torsional Alfv\'en wave does not inhibit the formation of small scales. On the contrary, the larger the turbulent component the more efficient the dissipation is. Indeed, as we shall see, the Alfv\'en wave is still subject to phase mixing even when it is distorted by a turbulent component. 

What is more, the fastest dissipation is found when the turbulent component dominates ($\alpha\simeq 1$). In that case the perturbation simultaneously undergoes phase mixing and nonlinear cascade. The combined effect of these two mechanisms leads to the most efficient dissipation. 
The case of Run 20, corresponding to $\alpha=1$ (Fig. \ref{1w2w}, upper panel), can be compared with that of a turbulent perturbation in a homogeneous background (Run 14), which is illustrated in Fig. \ref{j2turb}. Both those cases correspond to a perturbation with the same amplitude and the same hyper-dissipative coefficient, the only difference being in the background (inhomogeneous in Run 20 and homogeneous in Run 14). A comparison shows that the presence of inhomogeneity in the background structure strongly reduces the dissipative time. This clearly illustrates how phase mixing and nonlinear effects jointly act to promote dissipation.

A further effect that accelerates dissipation is related to modifications of the background density induced by the velocity field $\delta \bm{v}_\perp$ associated with fluctuations. In the case of a pure torsional wave it is $\delta \bm{v}_\perp=\delta v_\perp(r,z) \hat{\bm{\theta}}$; moreover, the background density is initially is independent of $\theta$:  $\rho_0=\rho_0(r)$. In this case it is $\nabla \cdot \left( \rho_0 \delta \bm{v}_\perp\right)=0$ and Eq. (\ref{MHD1}) implies $\partial \rho_0/\partial t=0$.
Therefore, velocity fluctuations do not modify the background density, and the cylindrical symmetry of the equilibrium is preserved. In contrast, when the turbulent component is present in the velocity fluctuations ($\alpha \ne 0$) the cylindrical symmetry is violated and the velocity $\delta \bm{v}_\perp$ can modify the background density structure. In spite of the small amplitude of fluctuations ($A=0.01$), this effect accumulates, and deformations in the density structure become more and more relevant with increasing time.

In Figure \ref{alpha2} we report a 3D visualization of the density field at times $t=200$ (top), $t=450$ (center), and $t=700$ (bottom), for the Run 20 where only the turbulent component is present in the fluctuations ($\alpha=1$). It can be seen how the initial cylindrical symmetry is gradually lost, and increasingly larger deformations are produced in the density until, at later times, the monolithic structure of the loop is completely destroyed. Of course, this effect is stronger when the turbulent component dominates fluctuations ($\alpha \sim 1$). Such effect is similar to as that reported in \citet{magyar2017generalized}, where density initial inhomogenities are deformed by wave propagation. 

Another feature visible in Fig. \ref{alpha2} is the gradual formation of small scales in the background density, transverse to $\bm{B}_0$, which is driven by small-scale generation in the velocity perturbation $\delta \bm{v}_\perp$.
Small scales in the density correspond to small scales in the spatial distribution of the Alfv\'en velocity $c_A(x,y)$. This leads to the formation of regions where the Alfv\'en velocity gradient is much larger than in the initial equilibrium. Since the phase-mixing dynamical time is proportional to $|\nabla c_A|^{-1}$ (Eq. \ref{tPM0}), in those regions, the generation of small scales proceeds with a rate larger than at the initial time. This effect, which becomes more relevant with increasing $\alpha$, contributes to speeding up the dissipation.

\begin{figure}
\hspace{-0.8cm}
\includegraphics[scale=0.35]{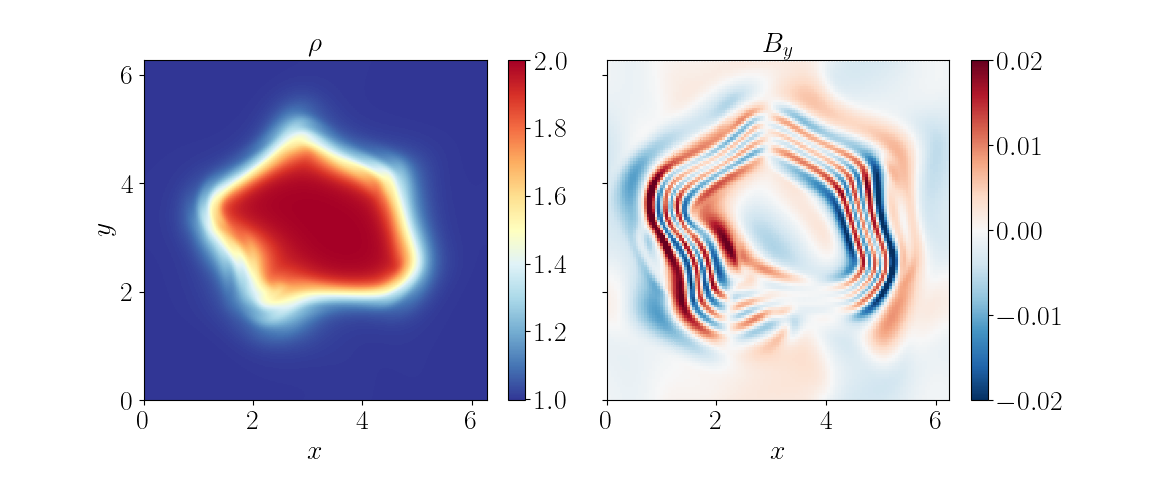}
\caption[...]
{\footnotesize{} 2D perpendicular section of the density field (left panel) and the $B_y$ component (right panel) for the Run 16 at $t\sim t_d = 300$ and $z=L_\parallel /2$. The simulation corresponds to a propagating Alfv\'en wave with a moderate level of turbulence superposed ($\alpha=0.23.$).}
\label{comp_pw}
\end{figure}

However, deformations in the density structure are also observed for lower values of $\alpha$, though smaller than for $\alpha=1$. 
We report a 2D section of the density field for the propagating wave (left panel of Fig. \ref{comp_pw}) at $z=L_\parallel /2$ and time $t=300$.
This Figure refers to Run 16, where we used the value $\alpha=0.23$. In this case, the percentage of turbulence in the initial perturbation is not enough to completely destroy the loop, but it is able to generate relevant changes in the morphology of the background density.
In the right panel of the same figure, we report the $B_y$ field component at the same time and at the same value of $z$.
The simulation illustrated in Fig. \ref{comp_pw} corresponds to a case of a torsional Alfv\'en wave plus a moderate level of turbulence, and can be then compared with Fig. \ref{alfvenevo} where the case $\alpha=0$ (Run 15) is illustrated. In Fig. \ref{alfvenevo} it is visible how phase mixing produces wavefronts in the form of perfect concentric shells. Differently, for $\alpha=0.23$ (Fig. \ref{comp_pw}) the cylindrical symmetry is lost; nevertheless, phase mixing still takes place, locally generating small scales in the form of closely packed wavefronts. The wave vector associated with those wavefronts is locally directed parallel to the background density gradient. This shows that, even in the presence of deformations in the density and wave patterns, phase mixing proceeds with properties similar to what found in the cylindrically symmetric configuration and that the presence of turbulence makes dissipation more efficient.

\subsubsection{Standing Alfv\'en wave plus turbulence}
\label{sec3.3.2}

A coronal loop with a finite length $L_{||}$ supports standing waves. Torsional Alfv\'en waves are among the possible standing modes. Therefore,
in addition to the previous ones, we also performed simulations of a standing torsional Alfv\'en wave (formed by two counter-propagating waves) with a variable amount of turbulence superposed on it.
In order to make a comparison, we adopted the same physical and numerical parameters used in the runs of the propagating waves, except for the amplitude of the initial magnetic perturbation that is greater by a factor $\sqrt{2}$. As explained in Sect. \ref{Aws}, this choice ensures that the energy of the standing wave is the same as the propagating wave. Different tests have been performed with various values of the parameter $\alpha$; those runs are indicated as Run 21--26 in Table \ref{table1}. We observe that the case $\alpha=1$ described in Run 26 is coincident with that of Run 20, since in both runs only the turbulent component is present. Nevertheless, we indicate this same run with two different numbers to more easily indicate how results vary when varying $\alpha$ in the propagating and standing wave cases.

The time evolution of the $W$ is illustrated in the bottom panel of figure \ref{1w2w} for this set of runs. Similar as in the propagating wave case, in the present situation the dissipative time decreases with increasing the relative turbulence level (increasing $\alpha$). At the same time, the maximum value of $W$ decreases with increasing $\alpha$. For $\alpha \gtrsim 0.66$ the time evolution of $W$ becomes essentially independent of $\alpha$. 
Comparing the two panels of Fig. \ref{1w2w}, it appears that the time behavior of the dissipation rate $W$ in the standing wave cases is qualitatively similar as for the propagating wave, though with some small differences.

\begin{figure}
\hspace{-0.8cm}
\includegraphics[scale=0.35]{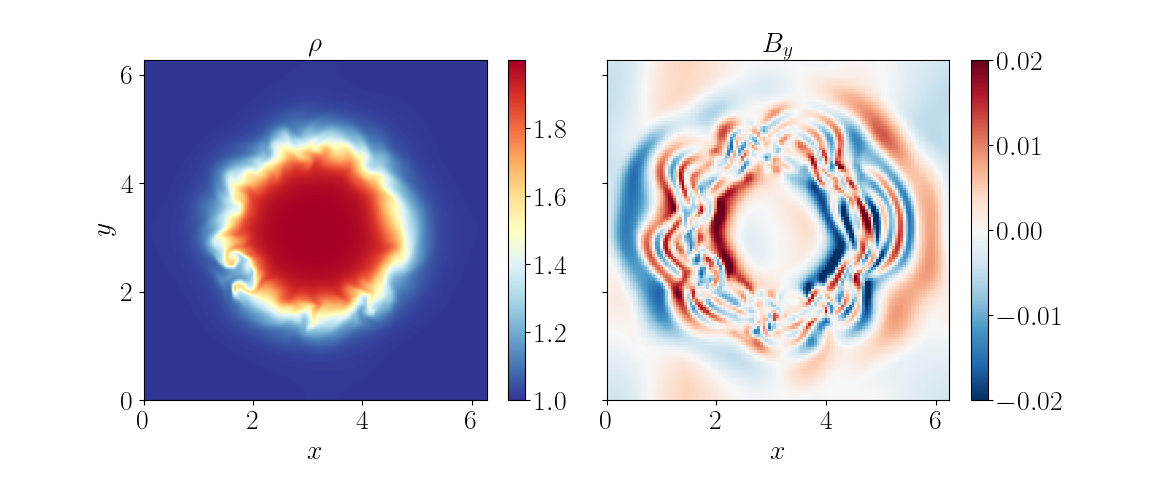}
\caption[...]
{\footnotesize{}  2D perpendicular section of the density field (left panel) and the $B_y$ component (right panel) for the Run 22 at $t\sim t_d = 300$ and $z=L_\parallel /2$. Here we evolve a standing Alfv\'en wave with a moderate level of turbulence superposed ($\alpha=0.23.$).}
\label{comp_sw}
\end{figure}

In Fig. \ref{comp_sw}, we report a 2D section illustrating the density profile (left panel) and the $B_y$ component of the magnetic field (right panel) at $z=L_\parallel /2$ and $t=300$, for Run 22 where a moderate amount of turbulence ($\alpha=0.23$) is initially present.
This can be compared with the corresponding Run 15 (Fig. \ref{comp_pw}) relative to a propagating wave case.
The comparison shows net differences in morphology. In particular, in the standing wave case, the flux tube boundary appears to be fringed with curly filaments. Such small-scale structures are generated by rolls related to the formation of a KH instability that develops at the loop boundary as a consequence of phase mixing of the standing Alfv\'en wave. A similar phenomenon also takes place when an initial standing kink mode couples with Alfv\'enic oscillations that undergo phase mixing \citep[see, e.g.,][]{Antolin16}. In our case, where a torsional Alfv\'enic standing wave is considered, a certain amount of turbulence ($\alpha \ne 0$) must be present; otherwise, the KH instability is not observed.

Similar as in the propagating wave case, we observe the formation of packed wavefronts in the $B_y$ magnetic field, located across the loop boundary. Such wavefronts appear to be distorted by the underlying irregular shape of the background density (left panel of the same figure). Comparing with the corresponding pattern obtained in Run 16 at the same time, we see that, in the propagating wave case, $B_y$ contains scales that are on average smaller than in the standing wave case. This is in accordance with the profiles of $W$ shown in Fig. \ref{1w2w} for $\alpha=0.23$ in the two cases. In fact, for $t\sim t_d$ the dissipation rate $W$ for the propagating wave is slightly larger than for the standing wave (orange lines).
In both cases, it can be seen that small scales generated by phase mixing are localized across the region of the density gradient, which corresponds to the location of the strongest Alfv\'en velocity gradient.

To complement the comparison between the cases of propagating and standing waves, we computed the perpendicular power spectra of the magnetic field transverse components, namely $P(k_\perp) = |\hat{B_x}(k_\perp)|^2 + |\hat{B_y}(k_\perp)|^2$, where the hat refers to the Fourier coefficients of the fields. 
Such spectra are calculated by integrating the 3D Fourier spectra both on concentric 2D shells located perpendicularly to the $k_z$ direction and along $k_z$. The result is a spectrum that depends only on $k_\perp$. 
In Fig. \ref{spec}, we report the magnetic field power spectra for Run 16 and for Run 22 as a function of $k_\perp$ and $t \sim t_d=300.$ A spectrum proportional to $k_\perp^{-5/3}$ (black dashed line) is plotted for reference. It can be seen that the two spectra at $\alpha=0.23$ are similar and both are less steep than a Kolmogorov spectrum $\propto k_\perp^{-5/3}$. Indeed, for such a moderate value of $\alpha$ the dynamics is dominated by phase mixing instead of by a turbulent cascade. While in a turbulence the amplitude of fluctuations decreases with increasing the wavevector, phase mixing increases the perturbation wavevector keeping the amplitude unchanged, at least in the region where the background inhomogeneity is located. As a result, the Fourier spectra calculated over the entire spatial domain for Runs 16 and 22 are shallower than a Kolmogorov spectrum. 
At the smallest scales ($k_\perp \gtrsim 30$), the spectrum of the standing wave case have more energy than for the propagating wave. However, these differences are at energies too small to be appreciated in the perturbation morphology.

To conclude, from the point of view of wave dissipation, there are no relevant differences between the cases of propagating and standing waves. In particular, the presence of KH rolls that characterize standing waves does not particularly enhance dissipation, at least in the case of a distorted torsional wave we are considering.

\begin{figure}
\hspace{-0.8cm}
\includegraphics[scale=0.35]{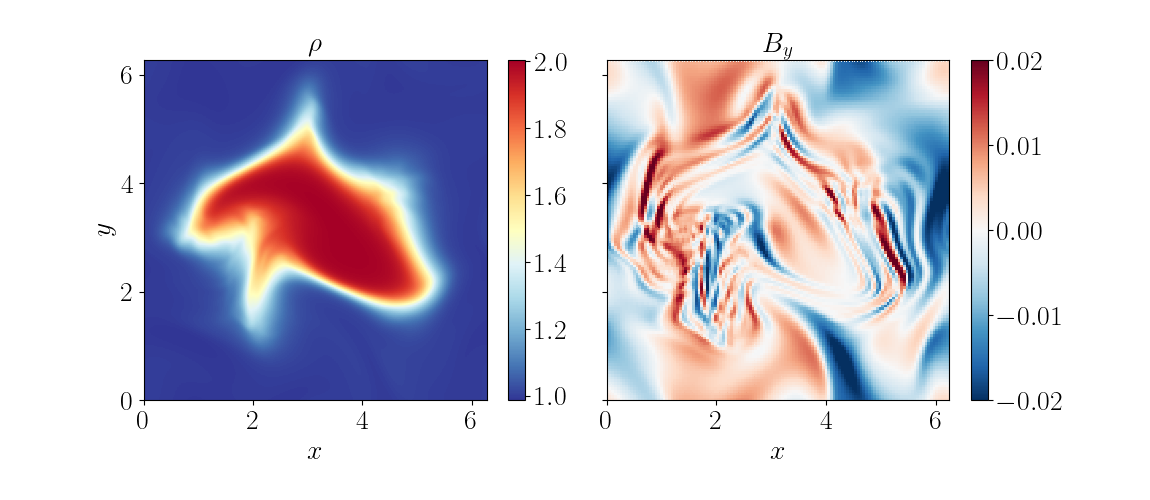}
\caption[...]
{\footnotesize{} $xy-$plane for the density field (left panel) and the $B_y$ component (right panel) for the Run 20 at $t \sim t_d = 200$ and $z=L_\parallel /2$. 
Here we used the maximum level of relative turbulence ($\alpha=1.0$). }
\label{comp_t}
\end{figure}

\subsection{Turbulence in inhomogeneous background}
\label{sec3.4}

The case where the initial perturbation is formed only by the turbulent component ($\alpha=1$) has been considered in Run 20. Results are illustrated in Fig. \ref{comp_t}, where the density $\rho$ and the $B_y$ component of the magnetic field are reported, at $t=t_d=200$, similar as in Fig.s \ref{comp_pw}-\ref{comp_sw}. 
Even in this case, the pattern of $B_y(x,y)$ is reminiscent of what is generated by phase mixing, namely, small-scale structures appear, formed by closely packed wavefronts. Those structures are essentially localized in regions of inhomogeneous density. Therefore, the coupling between the turbulent perturbation and the inhomogeneous background leads to a phenomenon that is qualitatively similar to phase mixing.

However, in Fig. \ref{comp_t} some regions are visible (see, e.g., the area $1\lesssim x \lesssim 3$, $1\lesssim y \lesssim 3$) where the wavevector distribution appears to be more isotropic, also in comparison with cases characterized by lower values of $\alpha$ (e.g., $\alpha=0.23$, Fig. \ref{comp_pw}). Those are the regions where the smallest scales are localized. This feature can be interpreted as a manifestation of a nonlinear cascade that forms locally following the development of phase mixing. 
We also observe that in those regions, the density structure is particularly distorted and presents a large gradient (Fig. \ref{comp_t}, left panel). This contributes to the formation of small scales in the perturbation. 

\begin{figure}
\includegraphics[scale=0.55]{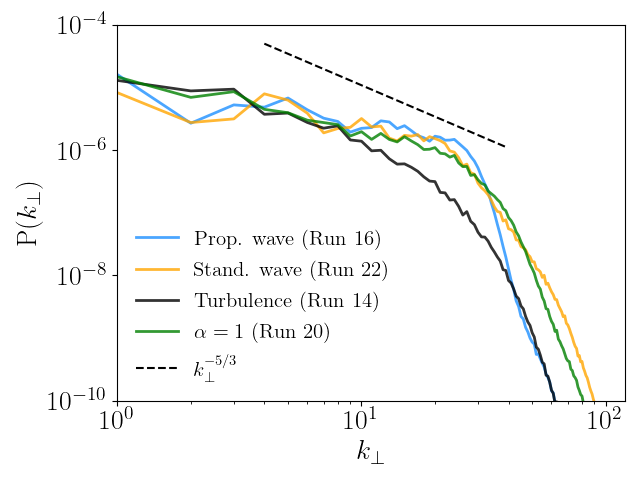}
\caption[...]
{\footnotesize{} $B_\perp^2(k_\perp)$ power spectra for different cases: the propagating Alfv\'en wave (Run 16), the standing Alfv\'en wave (Run 22),  the turbulence in homogeneous background (Run 14), and the turbulence in inhomogeneous background (Run 20). All the spectra are reported at the dissipative time of the corresponding run. We also report a $k^{-5/3}$ slope for comparison.}
\label{spec}
\end{figure}

In Fig. \ref{spec} the spectrum of magnetic fluctuations $P(k_\perp)$ is plotted for the case of turbulence in the flux tube (Run 20, green curve), as well as in the case of turbulence in an homogeneous background (Run 14, black curve). All the spectra plotted in Fig. \ref{spec} are calculated at the dissipative time of the corresponding run, namely the time at which $W$ is maximum, in order to have a spectrum as developed as possible. In the case of Run 14 we observe the presence of an inertial range that approximately follows the slope $\propto k^{-5/3}$, indicating that the spectrum has been formed by the nonlinear cascade. In contrast, in the case of turbulence in the flux tube (Run 20) a shallower spectrum is observed that is close to the spectra of phase-mixing-dominated Runs 16 and 22. This further demonstrate that, when the turbulence evolve in the inhomogeneous flux tube, phase-mixing plays a relevant role in the generation of small scales and dissipation. At small scales $k_\perp > 10-20$ the spectrum of Run 20 becomes less steep than those of Runs 16 and 22, giving origin to a more efficient dissipation. This feature could be interpreted as a sign of other phenomena (nonlinear cascade, deformation of the background density) that cooperate with phase mixing in the generation of small scales.

\section{Summary and conclusions}
\label{sec4}

In the present paper, we have investigated the interplay between phase mixing and turbulence in generating small scales and consequent wave dissipation in a simple model of a coronal loop. Both effects tend to generate small scales transverse to the background magnetic field in Alfv\'enic fluctuations, and therefore, they can jointly act to promote the dissipation of fluctuations. We have considered a configuration where an initial torsional Alfv\'en wave and a variable amount of turbulent Alfv\'enic transverse fluctuations are present. We chose a torsional wave because its time evolution is essentially determined by phase mixing, which is one of the two mechanisms we are interested in investigating. There is observational evidence of torsional motions, both in the photosphere \citep{Brant88,Bonet08,Bonet10} and in the chromosphere \citep{Wedemeyer09,Wedemeyer12,Tziotziou18,Dakanalis22}, which could induce torsional waves in coronal loops \citep{Jess09}.

In this study, we used the \texttt{COHMPA} algorithm \citep{pezzi2023turbulence}, which solves the 3D compressible MHD equations with periodic boundary conditions. We considered an equilibrium structure with a low $\beta$ value, representing an overdense cylindrical magnetic flux tube. The Alfv\'en speed gradient across the flux tube boundary is responsible for the phase mixing of Alfv\'enic perturbations. We characterized the time evolution of a pure torsional Alfv\'en wave by varying parameters like wave amplitude and parallel wavelength. We verified that the properties of the dissipation rate in the numerical runs follow what can be predicted for the phase mixing to a good extent. However, the dynamical time associated with phase mixing increases with decreasing the spatial scale $\ell_\perp$ of perturbations, and it becomes exceedingly long for realistic values of the dissipative scale in the coronal plasma.
In this respect, the turbulent cascade has the advantage of moving the fluctuation energy down to small dissipative scales in a time that is of the order of the large-scale nonlinear time. For that reason MHD turbulence has been considered in several coronal heating models \citep{Nigro04,Malara10,vanBallegooijen17,rappazzo2017,vanBallegooijen18}. In addition, there are indications of the presence of turbulent oscillations in the corona \citep{Banerjee98,Singh06,Hahn13,Hahn14,Morton16,Morton19}. 

A torsional wave propagating in a cylindrical flux tube represents a peculiar configuration that is completely independent of the azimuthal coordinate $\theta$. It is natural to ask to what extent the phenomenon of phase mixing is related to such a particular symmetry and whether phase mixing is still present when the cylindrical symmetry is violated. In our simulations, we investigated this question by distorting the pure torsional wave with the addition of a certain amount of turbulent perturbation. Our results have shown that, for moderate amplitudes of the turbulent component (small $\alpha$), phase mixing persists even when the cylindrical symmetry is not preserved: closely packed wave fronts form, with a wave vector that is locally parallel to the density gradient in planes perpendicular to $\bm{B}_0$ (Fig. \ref{comp_pw}). In this respect, phase mixing appears to be a robust mechanism active in configurations more complex than what originally conceived \citep{Heyvaerts83}. 

We have also considered the case of standing Alfv\'en wave with a variable amount of turbulence superposed on it. In this case, the evolution of the background structure suggests that rolls form at the flux tube boundary, probably generated by a mechanism similar to a KH instability. This phenomenon has been previously observed in another situation, where an initial kink mode of an overdense flux tube couples with azimuthal Alfv\'enic perturbations which, in turn, undergo phase mixing \citep[e.g.,][]{Antolin16}. In our case, where the initial perturbation is a torsional Alfv\'en wave, this phenomenon requires a certain amount of turbulent distortion in the initial perturbation ($\alpha \ne 0$), otherwise the cylindrical symmetry prevents the formation of the KH instability. Moreover, KH rolls are not observed in the case of the propagating wave, even in the presence of a turbulent component. The velocity perturbation pattern, with its crests and troughs, must probably remain stationary to allow for the development of the KH instability. However, comparing the propagating and standing waves, we did not find relevant differences in the dissipative time for the two cases (Fig. \ref{1w2w}). Therefore, the presence of KH rolls seems not to enhance the efficiency of small-scale formation, at least for the considered configuration. 

Increasing the parameter $\alpha$ up to $\alpha=1$ the initial perturbation becomes dominated by the turbulent component. The case $\alpha=1$ corresponds to the shortest dissipative time. Even in such a case, where no torsional wave is initially present, the time evolution of the turbulent perturbation reveals features qualitatively similar to what is produced by phase mixing. Namely, closely packed wavefronts localized at the flux tube boundary, with a wave vector locally parallel to the density gradient (Fig. \ref{comp_t}). This indicates that a mechanism similar to phase mixing acts -- driven by transverse inhomogeneities in the Alfv\'en velocity -- even on a turbulent Alfv\'enic perturbation. 

The case of a turbulent perturbation evolving in a flux tube has been studied in a regime expressed by Eq. (\ref{cond}). 
In this case, the amplitude of the initial turbulent perturbation is small enough to make the nonlinear time -- calculated for the large, energy-containing scales -- longer than the corresponding phase-mixing dynamical time. 
In these conditions, the small-scale formation within the turbulent perturbation is initially dominated by phase mixing rather than by the nonlinear cascade. However, while phase mixing proceeds, the perturbation energy moves toward small scales, thus reducing the effective nonlinear time. Eventually, the nonlinear cascade becomes faster than phase mixing and builds up small scales in a time that is essentially independent of the dissipative coefficients. This aspect is relevant for the coronal plasma, where such coefficients are expected to be very small. 
In the considered regime, our simulations have shown that the corresponding dissipative time is shorter than what is found both in a pure phase mixing case (represented by the case $\alpha=0$) and for a standard turbulent cascade (Run 14). Therefore, phase mixing and turbulence work in a synergistic way, speeding up the dissipation of the perturbation energy.

\begin{figure}
\includegraphics[scale=0.7]{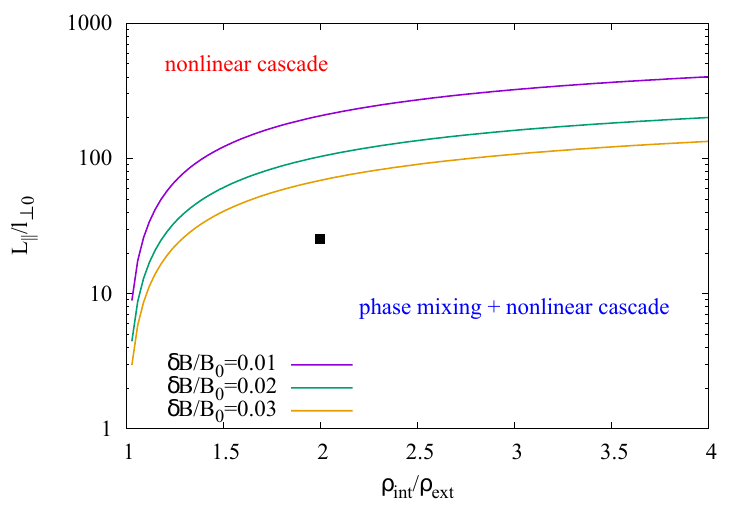}
\caption[...]
{\footnotesize{} Curves $F( \rho_{int}/\rho_{ext},L_{||}/\ell_{\perp 0}, \delta B/B_0, l_{\perp 0}/\Delta r)=0$ are represented in the $(\rho_{int}/\rho_{ext}, L_{||}/\ell_{\perp 0})$-plane, for $\delta B/B_0=0.01$ (purple line), $\delta B/B_0=0.02$ (green line), and $\delta B/B_0=0.03$ (orange line); each curve is calculated for $\ell_{\perp 0}/\Delta r=1$. The portion of the plane below a given curve corresponds to the regime defined by the condition (\ref{cond}) ("phase mixing + nonlinear cascade" regime), while the portion above the curve corresponds to the opposite condition ("nonlinear cascade" regime). The black dot indicates the configuration considered in Run 20.}
\label{fig_cond}
\end{figure}

We refer to the above situation as a "phase mixing + nonlinear cascade" regime. The opposite case corresponds to a perturbation amplitude large enough to violate the condition (\ref{cond}). In this case, the nonlinear time is shorter at large scales than the phase-mixing dynamical time. Therefore, the nonlinear cascade dominates the generation of small scales, with phase mixing playing a minor role. We refer to this latter situation as a "nonlinear cascade" regime. 
To better illustrate these two regimes, in the condition (\ref{cond}), we express the gradient of the Alfv\'en velocity as $dc_A/dr \sim \Delta c_A/\Delta r$, where $\Delta c_A=c_{A,ext}-c_{A,int}$ is the variation of the Alfv\'en velocity across the flux tube boundary, which has a width $\Delta r$, and $c_{A,ext}=B_{ext}/\sqrt{4\pi \rho_{ext}}$ and $c_{A,int}=B_{int}/\sqrt{4\pi \rho_{int}}$. Similarly, we estimate the Alfv\'en velocity as $c_A \sim (c_{A,ext}+c_{A,int})/2$. For small values of the plasma $\beta$, we can assume that $B_{ext}\simeq B_{int}\simeq B_0$. Therefore, it is
\begin{equation}\label{Alfratio}
\frac{dc_A}{c_A}\sim 2\frac{\sqrt{\rho_{int}/\rho_{ext}}-1}{\sqrt{\rho_{int}/\rho_{ext}}+1}
\end{equation}
We can re-express the the condition (\ref{cond}) in terms of the dimensionless ratios $\rho_{int}/\rho_{ext}$, $L_{||}/\ell_{\perp 0}$, $\ell_{\perp 0}/\Delta r$ and $\delta B/B_0$ characterizing the problem, in the following form:
\begin{equation}\label{cond2}
F\left( \frac{\rho_{int}}{\rho_{ext}},\frac{L_{||}}{\ell_{\perp 0}}, \frac{\delta B}{B_0}, \frac{l_{\perp 0}}{\Delta r} \right) \equiv
\frac{L_{||}}{\ell_{\perp 0}} -
2\Gamma \frac{\left( \sqrt{\displaystyle{\frac{\rho_{int}}{\rho_{ext}}}}-1\right)}{\left( \sqrt{\displaystyle{\frac{\rho_{int}}{\rho_{ext}}}}+1\right)} \,
\frac{\left( \displaystyle{\frac{\ell_{\perp 0}}{\Delta r}}\right)}{\left(\displaystyle{\frac{\delta B}{B_0}}\right)} < 0
\end{equation}
The condition (\ref{cond2}) is represented in Fig. \ref{fig_cond} where lines corresponding to $F=0$ are plotted in the $(\rho_{int}/\rho_{ext}, L_{||}/\ell_{\perp 0})$-plane, for three values of the ratio $\delta B/B_0$ ($\delta B/B_0=0.01,0.02,0.03$), and choosing $\ell_{\perp 0}/\Delta r=1$ (as in our simulations). The portion of the plane below each curve corresponds to configurations satisfying the condition (\ref{cond2}) (or, equivalently, condition (\ref{cond})). Therefore, it corresponds to the regime denoted as "phase mixing + nonlinear cascade". This is verified for high values of the density contrast $\rho_{int}/\rho_{ext}$ and for a longitudinal scale $L_{||}$ that is not exceedingly larger than the perpendicular scale $\ell_{\perp 0}$. The configuration considered in Run 20 is indicated by a black dot, which falls into this regime. The opposite regime ("nonlinear cascade") corresponds to low-density contrast and/or high values for the ratio $L_{||}/\ell_{\perp 0}$. In this case, phase mixing is slower than nonlinear couplings, and the process of small-scale generation is dominated by the nonlinear cascade. Since the nonlinear time scale decreases with increasing the perturbation amplitude, the portion of the plane corresponding to this latter regime becomes larger when the ratio $\delta B/B_0$ is increased. The above considerations indicate that the effects of phase mixing in the dynamics and dissipation of turbulent perturbations are more relevant in coronal loops with a high-density contrast and short length. 

Despite the small amplitude of the considered perturbations, their effects on the background structures are not negligible when considered for a sufficiently long time. In particular, the initial density structure is gradually distorted, losing its cylindrical symmetry. This effect becomes more relevant at large $\alpha$, and it can completely destroy the initial structure of the flux tube after a certain time. This leads to the formation of regions where the gradient of the Alfv\'en velocity is particularly large. In those regions, the generation of small scales in the perturbation proceeds with a rate higher than elsewhere (Fig. \ref{comp_t}). This effect further contributes to accelerate dissipation. Therefore, it can be expected that such distortion of the background structure in a coronal loop corresponds to locations of greater heating. 

In previous work, it has been shown that multiple density patches coupled with a large scale wave eventually result in a turbulent like distribution of the density \cite{magyar2017generalized,magyar2019understanding}. Here, we have investigated the case of a single inhomogeneity perturbed by a spectrum fluctuations at large scale, focusing our interest on the interplay between generalized phase mixing (following the definition by \citet{magyar2017generalized}) and nonlinear effects.  

In summary, phase mixing appears to be a robust mechanism that works both in the less idealized case of a distorted torsional wave and on a fully turbulent perturbation. We conclude that phase mixing, turbulent cascade, and perturbation-induced background distortions jointly work to promote wave dissipation. The present results are relevant for the solar corona, where the plasma is characterized by extremely high Reynolds and Lundquist numbers' values. In that ambient, continuous injection of fluctuations from photospheric motions is expected, which demands forced rather than decaying simulations as those presented in this paper. Such a study is left for future work.

\begin{acknowledgements}
The present work has been supported by the Italian Space Agency (ASI) within the Project "Partecipazione italiana alla missione NASA/MUSE" (CUP: F83C22001920005), related to the NASA mission "Multi Slit Solar Explorer (MUSE)".
The simulations have been performed at the Newton cluster at University of Calabria and the work is supported by “Progetto STAR 2-PIR01 00008” (Italian Ministry of University and Research). SS acknowledges supercomputing resources and support from ICSC–Centro Nazionale di Ricerca in High-Performance Computing, Big Data and Quantum Computing–and hosting entity, funded by European Union–NextGenerationEU.
\end{acknowledgements}

\bibliographystyle{aa}
\bibliography{bibliografia}

\end{document}